\documentclass[aps,prl,superscriptaddress,showpacs,floatfix]{revtex4-1}
\usepackage{graphicx,amsmath,amssymb,xspace,epsfig,float,multirow,subfigure,tabularx}
\usepackage{amsbsy,amssymb,amsmath,bm}
\usepackage{epsf}
\usepackage{color,natbib}
\usepackage{amsfonts}
\usepackage{hyperref}

\pdfoutput=1

\begin{document}

\title{Soft phonons in the interface layer of the STO substrate can explain
high temperature superconductivity in one unit cell $FeSe$.}
\author{Baruch Rosenstein}
\email{vortexbar@yahoo.com}
\author{B. Ya. Shapiro}
\email{shapib@biu.ac.il}

\begin{abstract}
Using a microscopic model of lattice vibrations in the STO$\left( 001\right) 
$ substrate, an additional $\Omega _{s}=50mev$ longitudinal optical (LO)
interface mode is identified. The soft mode propagating mainly in the first $%
TiO_{2}$ layer ("$O$ chains") has stronger electron - phonon coupling to
electron gas in $FeSe$ than a well known $\Omega _{h}=100mev$ hard mode. The
coupling constant, critical temperature, replica band are calculated.
Although there exists a forward in the electron - phonon scattering peak, it
is clearly not as sharp as assumed in recent theories (delta function -
like). The critical temperature is obtained by solution of the gap equation
and agrees with the observed one. The corresponding electron phonon coupling
constant $\lambda =0.23$. The quasiparticle normal state "satellite" in
spectral weight is broad and its peak appears at frequency much higher than $%
\Omega _{s}$ consistent with observations usually associated with $\Omega
_{h}$. Possible relation of the transversal counterpart of the surface LO
soft mode with known phonons is discussed.
\end{abstract}

\maketitle

\affiliation{Electrophysics Department, National Chiao Tung University, Hsinchu 30050, 
\textit{Taiwan, R. O. C}} 
\affiliation{Physics Department, Bar-Ilan
University, 52900 Ramat-Gan, Israel}


\section{Introduction.}

\textit{\ }The best known group of superconductors with critical temperature
above $T_{c}=60K$, cuprates like $YBa_{2}Cu_{3}O_{7-\delta }$ ($93K$ at
optimal doping) and $Bi_{2}Sr_{2}CaCu_{2}O_{8+x}$ ($92K$), are generally
characterized by the following three structural/chemical peculiarities.
First they are all quasi - 2D perovskite layered oxides. Second the 2D
electron gas (2DEG) is created by maximally charging $CuO$ planes at optimal
doping. Superconductivity resides in these layers. Third the layers (or by
layers) are separated by several insulating ionic oxides. It is widely
believed\cite{Dagotto1} that, although the insulating layers play a role in
charging the $CuO$ planes, the bosons responsible for the pairing are
confined to the $CuO$ layer only.

Several years ago another group of high $T_{c}$ materials ($T_{c}=60-106K$)
was fabricated by deposition of a single unit cell (1UC) layer of $FeSe$ on
insulating substrates like $SrTiO_{3}$ (STO both\cite{expFeSe} $\left(
001\right) $ and\cite{110} $\left( 110\right) $), $TiO_{2}$ (rutile\cite%
{rutileFeSe} $\left( 100\right) $ and anatase\cite{anataseFeSe} $\left(
001\right) $) and\cite{Ba} $BaTiO_{3}$. It is interesting to note that all
three above features are manifest in this compounds as well. Indeed, the
insulating substrates are again layered perovskite oxides. The electron gas
residing in the $FeSe$ layer\cite{charging} that is charged (doped) by the
perovskite substrate. Of course there is a structural difference in that the
the layered cuprates contain many $CuO$ planes, while there is a single $%
FeSe $ layer. The difference turns out not to be that important, since
recently it was demonstrated\cite{XueBSCCO} that even a single unit cell $%
CuO $ on top of $Bi_{2}Sr_{2}CaCu_{2}O_{8+x}$ film still retains high $T_{c}$%
. Moreover the pairing becomes of the nodeless s-wave variety as in the
pnictides.

The role of the insulating substrate in the $FeSe$ systems however seems to
extend beyond the charging \cite{charging}. Although the physical nature of
the pairing boson in cuprates is still under discussion (the prevailing
hypothesis being that it "unconventional", namely not to be phonon -
mediated), it became clear that superconductivity mechanism in $FeSe$ should
at least include the substrate phonon exchange. There are several competing
theories. One is an unconventional boson exchange within the pnictide plane
(perhaps magnons \cite{Leerev}, like that in other pnictides.'
superconductivity theories\cite{Dagotto2}). It intends to explain both the $%
40K\,$(upon optimal charging) superconductivity in $K$ or $Li$ intercalated $%
FeSe$\cite{intercalatedFeSe} and "boosting" of superconductivity by an
interface $STO$ phonons above $60K$. Another point of view\cite%
{our,Gorkov,DFT16,JohnsonNJP16,Kulic} is that the "intrinsic" pairing in the 
$FeSe$ plane is dominated by the pairing due to vibration of oxygen atoms in
substrate oxide layers near the interface. Historically a smoking gun for
the relevance of the phonon exchange to superconductivity has been the
isotope effect. Very recently\cite{isotopeGuo} the isotope $^{16}O$ was
substituted, at least in surface layers of the $STO$(001) substrate, by $%
^{18}O$. For the same doping the gap at low temperature ($6K$) decreased by
about 10\%. Therefore the oxygen vibrations in the interface layers at least
influence superconductivity.

Moreover detailed measurements of the phonon spectrum via high resolution
electron energy loss spectroscopy (HREELS)\cite{Xue16phonon} were performed.
It demonstrated that the interface phonons are energetic ("hard" up to $%
\Omega _{h}=100mev$) for surface mode. This was corroborated by the DFT
calculations\cite{DFT16}. The phonons couple effectively to the electron
gas, as became evident from clear identification by ARPES of the replica band%
\cite{Lee12,Johnson16}. The explanation of the replica bands was based on
the forward peak in the electron - phonon scattering. Initially this
inspired an idea that the surface phonons alone could provide a sufficiently
strong pairing\cite{JohnsonNJP16,Kulic}. The values of the coupling constant
deduced from the intensity of the replica bands however was found to be
rather small $\lambda <0.25$. The BCS scenario, $T_{c}\approx \Omega
_{h}e^{-1/\lambda }$, is clearly out, even when possible violation of the
Migdal theorem due to nonadiabaticity ($\Omega _{h}>E_{F}$) is accounted for%
\cite{our}. One therefore had to look for other ideas. One is provided by a
possibility of the extreme, delta like, scattering peak model\cite{Kulicrev}%
, for which $T_{c}\approx \frac{\lambda }{2+3\lambda }\Omega _{h}$. Indeed
one can obtain\cite{JohnsonNJP16,Kulic} high $T_{c}$ even for such a small $%
\lambda $, but only for rather restrictive values of parameters of the ionic
substrate model (within the macroscopic dipole approximation
electrodynamics). Recently attempts were made to solve the Eliashberg
equation for the phonon mediated coupling\cite{Aperis} derived directly in
the framework of the density functional (DFT) approach\cite{DFTnew}.

In the present paper we consider a sufficiently precise microscopic model of
phonons in the ionic STO$\left( 001\right) $ substrate (beyond the
phenomenological dipole approximation approach) and find an additional much
softer LO interface mode that is as strongly coupled to the electron gas in
the $FeSe$ layer as the hard mode. The only parameters entering the model
are the Born-Meyer inter - atomic potentials\cite{Abrahamson} and measured
atomic charges\cite{averestov}. The coupling $\lambda $, critical
temperature, replica band and other characteristics of the superconducting
state are calculated and are consistent with experimental observations.
Although there exists a forward in the electron - phonon scattering peak, it
is clearly not a delta - like. The gap equations for the phonon - mediated
pairing are solved without this assumption. The soft mode $\Omega _{s}$
propagating mainly in the first $TiO_{2}$ layer ("$O$ chains") contribute
much more than the highest frequency $\Omega _{h}$ mode to the pairing.

\section{The interface structure, symmetry.}

\subsection{Structure of several top layers.}

The structure of the best studied high $T_{c}$ monolayer $FeSe$ system, that
on the STO substrate oriented along the $\left( 001\right) $ is as follows
The top three layers, where 2D electron gas resides, are $Se,Fe,Se$, while
the first substrate layer is $TiO_{2}$. The next layer is - $SrO$

Let us summarize experimentally determined configurations of atoms in the
one unit cell $FeSe/STO$ in a form sufficiently accurate for the phonon
spectrum calculation. The top three layers, $Se$ (1 and 3, green rings) and $%
Fe$ (brown ring), where 2D electron gas resides, are shown on the left of
Fig.SM1. The first substrate layer is $TiO_{2}$, as determined by STS is
shown in the center ($Ti$ - blue rings, $O$ - red full circle), while the
next layer is - $SrO$ (on the right, $Sr$ - cyan rings, $O$ - dark red full
circle). Below this plane the STO pattern is replicated. Out of plane
spacings counted from the $TiO_{2}$ layer are specified in Table I.

\begin{figure}[h]
\centering \includegraphics[width=6cm]{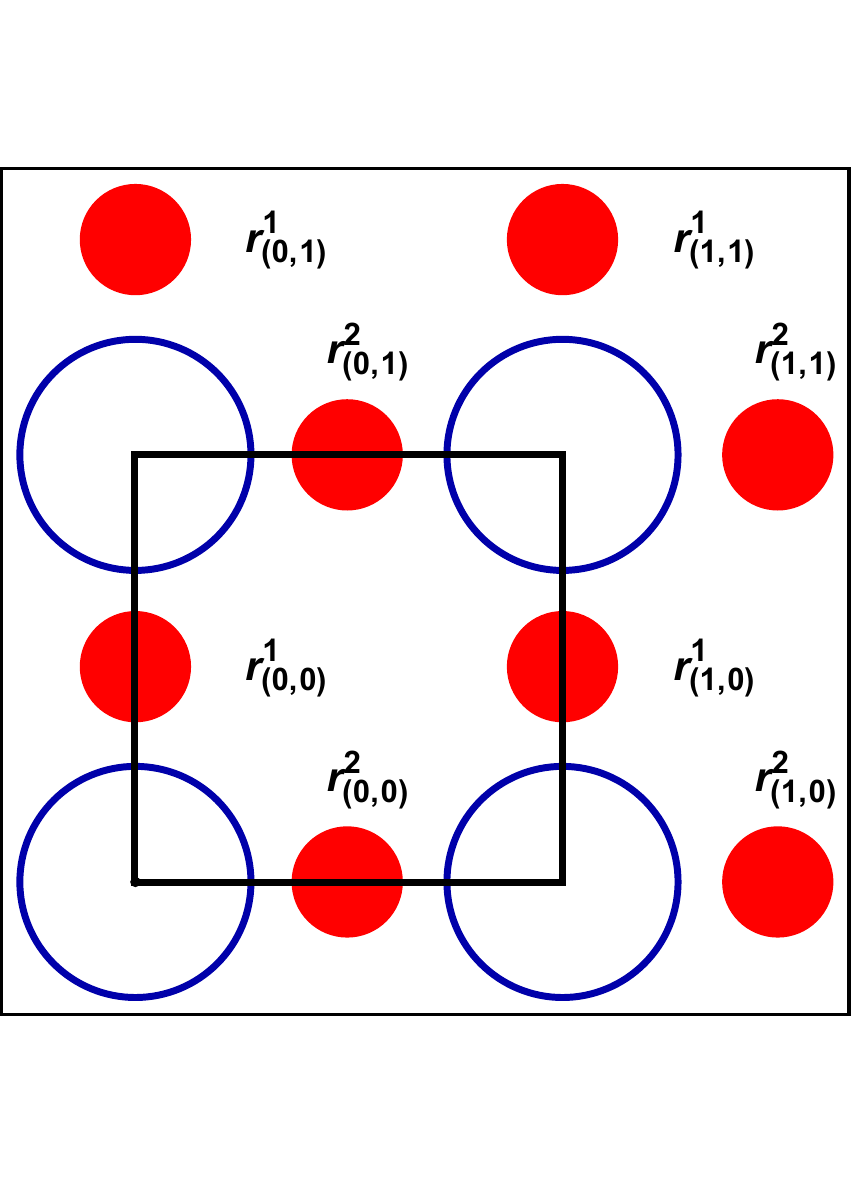}
\caption{ The top $TiO_{2}$ substrate layer. Blue empty circles represent
the $Ti$ atoms, while red filled ones represent light $O$ atoms. Unit cell
is depicted as a black square. Locations of the two sublattices in Eq.(2)
are shown. }
\end{figure}

\begin{figure}[h]
\centering \includegraphics[width=16cm]{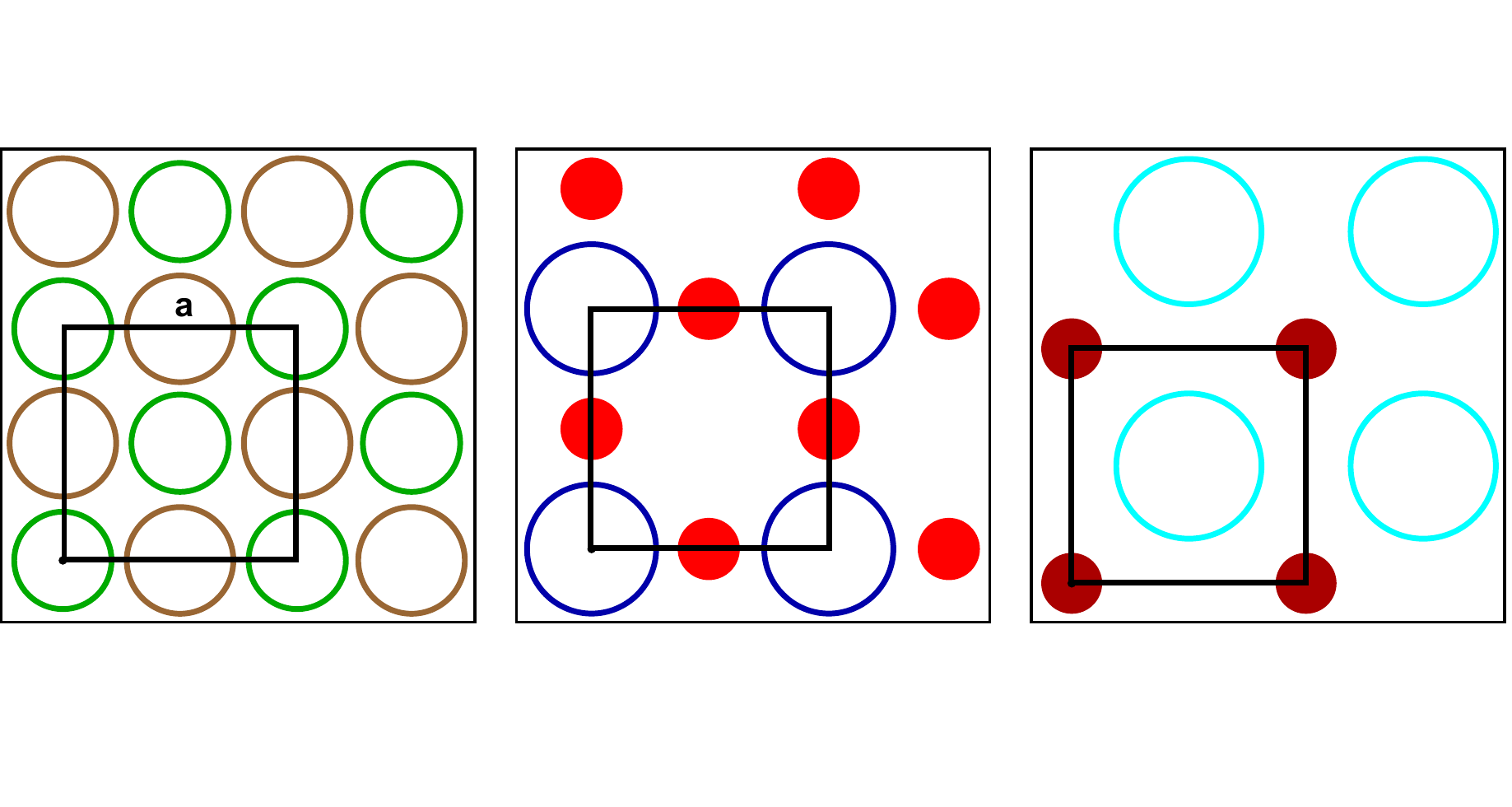}
\caption{ Structure of the surface layers. On the left $Fe$ (brown) and $Se$
(green) planes are projected. In the center the $TiO_{2}$ ($Ti$ - blue , $O$
- red), while on the right s the adjuscent $SrO$ ($St$ cyan, , $O$ - dark
red).}
\end{figure}
\begin{figure}[h]
\centering \includegraphics[width=7cm]{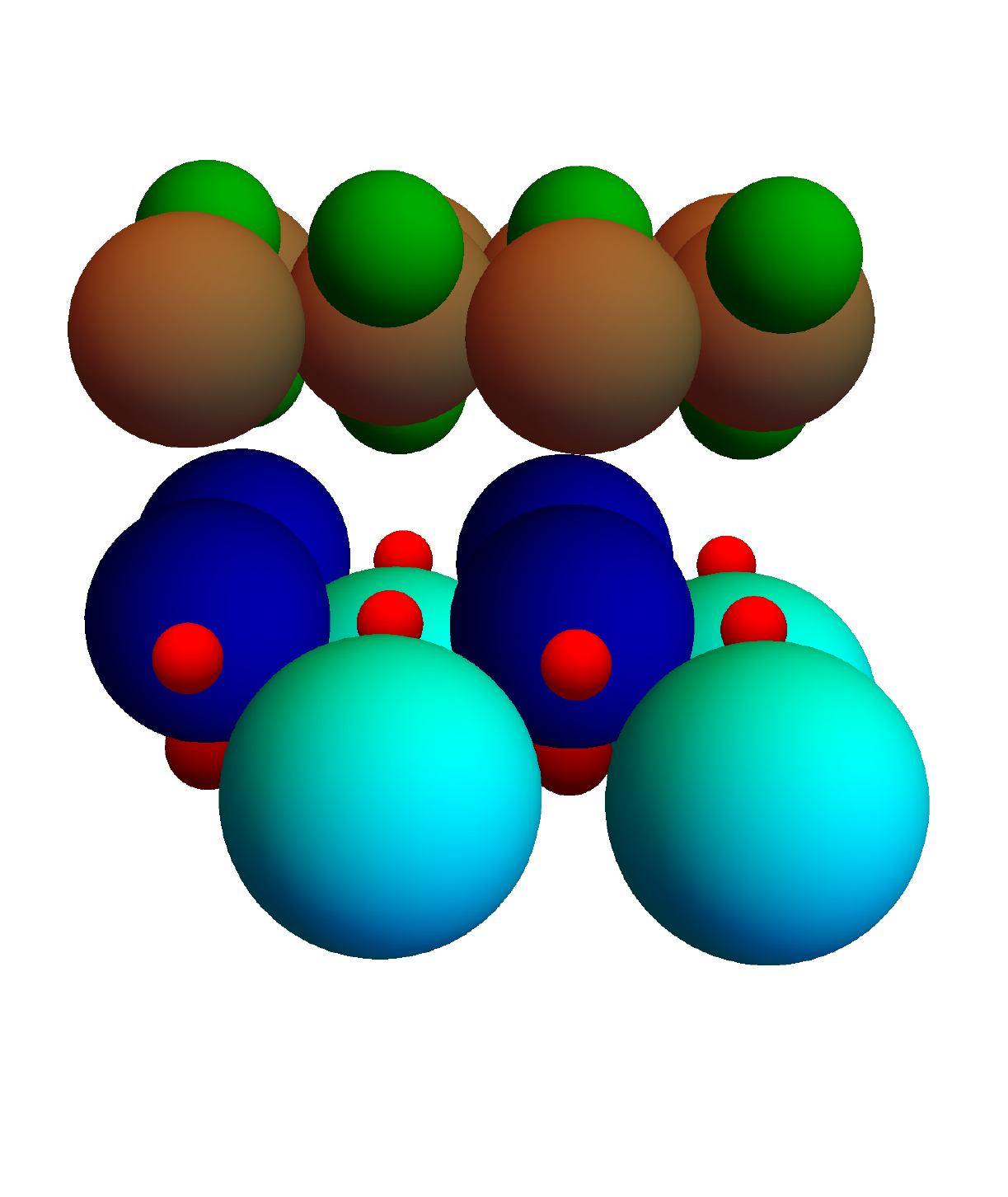}
\caption{3D view of the surface layers composing one unit cell :molecule".
Colors as in Fig.SM1. Sizes of atoms are inversely to the values of the Born
- Mayer parameter $b$.}
\end{figure}

Fig.3 is a 3D view of the molecule with sphere radii corresponding to the
repulsive Meyer potential ranges given in Table I.

\begin{table*}[h]
\caption{Atomic parameters.}
\begin{center}
\begin{tabular}{|c|c|c|c|c|c|}
\hline
$\text{atom}$ & $O$ in $TiO_{2}$ & $Ti$ & $Sr$ & $Se$ & $O$ in $SrO$ \\ 
\hline
$\text{mass (a.u.)}$ & $16$ & $48$ & $88$ & $79$ & $16$ \\ \hline
$A\, \  \text{(}kev\text{)}$ & $2.143$ & $9.353$ & $20.785$ & $17.56$ & $%
2.143 $ \\ \hline
$b$ \ ($A^{-1}$) & $3.388$ & $3.598$ & $3.541$ & $3.511$ & $3.388$ \\ \hline
charge $Z$ & $-1.27$ & $2.54$ & $1.85$ & $0$ & $-1.36$ \\ \hline
spacing $z$ ($A\,$) & $0$ & $0$ & $-1.6$ & $3.7$ & $-1.6$ \\ \hline
\end{tabular}%
\end{center}
\end{table*}

\subsection{The unit cell and symmetry of the whole system}

The square translational symmetry in the lateral ($x,y$) directions of the
system has two basis vectors shown in Fig.1, Unit cell including both the
metallic layer and the substrate containing $Fe_{2}Se_{2}TiSrO_{3}$ is
marked by the black frame in Fig. 4. The lattice spacing, that coincides
with the distance between the $Ti$ atoms is $a=3.9A$, equal to the distance
between the $Se$ atoms\cite{XueSTS}. The square translational symmetry in
the lateral ($x,y$) directions of the system has two basis vectors shown in
Fig.1. The lattice spacing, that coincides with the distance between the $Ti$
atoms is $a=3.9A$, equal to the distance between the $Se$ atoms\cite{XueSTS}.

\begin{figure}[h]
\centering \includegraphics[width=16cm]{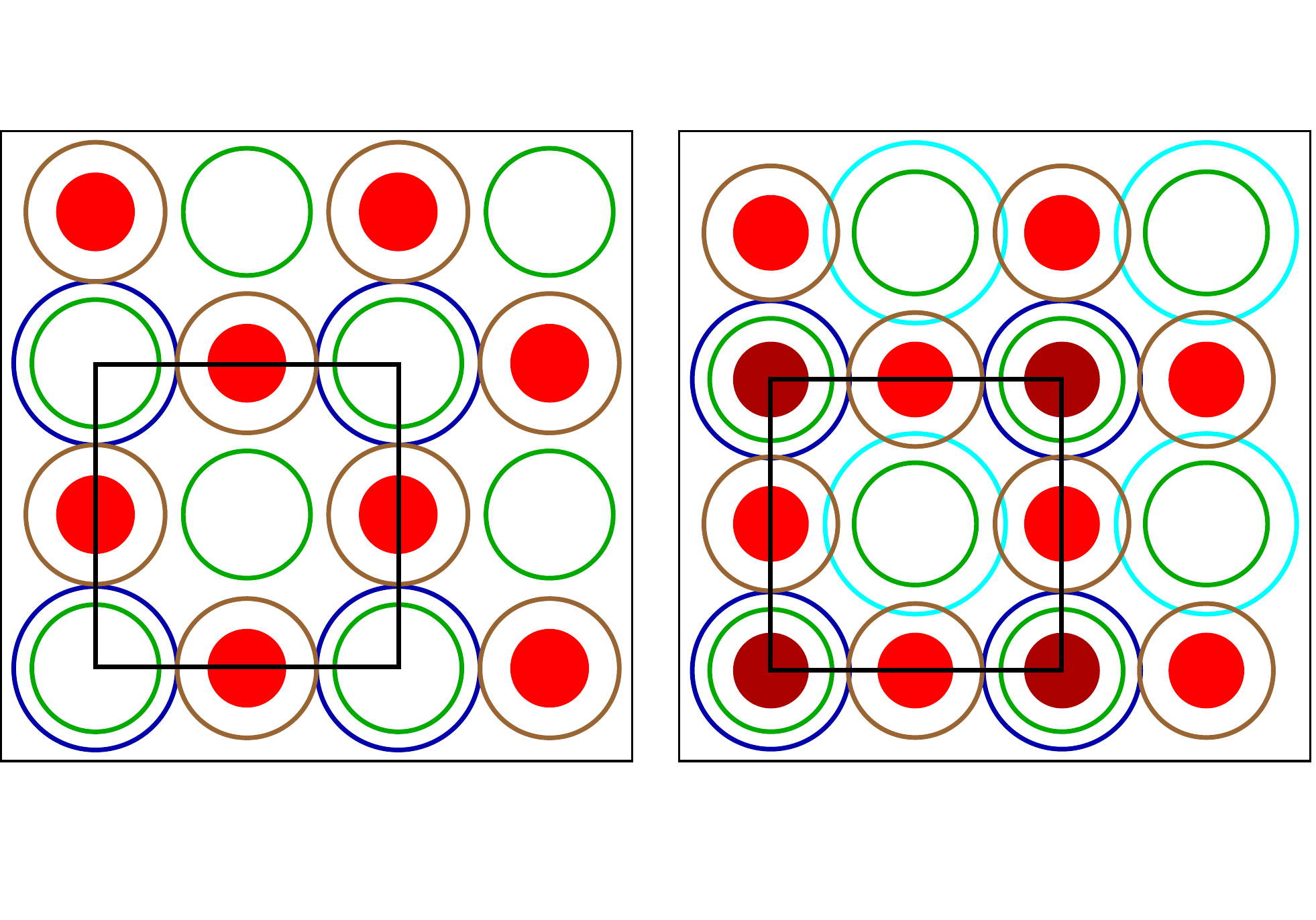}
\caption{Left. $FeSe$ with the first $TiO_{2}$ layer. Colors are same as in
Fig. SM1. }
\end{figure}

The left panel of Fig. 4 contains the projection of $FeSe$ and $TiO_{2}$,
while the whole "molecule" including the $SrO$ layer is given on the right
panel.

\section{Model the 2D electron gas in FeSe interacting with phonons in the
STO substrate.}

\subsection{Electron gas.}

Our model consists of the 2DEG interacting with surface phonons of polar
insulator $STO$:%
\begin{equation}
H=H_{e}+H_{ph}+H_{e-ph}\text{.}  \label{Hamiltoniandef}
\end{equation}

The Fermi surface consists of two slightly distinct electron pockets
centered around the crystallographic $M$ - point. Although experiment\cite%
{isotopeGuo} shows four - fold symmetry breaking, it is much smaller than
the asymmetry of the superconducting order parameter and will be neglected.
The electron gas is described sufficiently well by a simple tight binding
model on square lattice with spacing $a=3.9A$, proposed in ref \cite{DFT13}.
Electrons are hoping between the $Fe$ $4d_{xz}$ and $4d_{yz}$ orbitals
around locations of $Fe$ atoms on two sublattices, $A=1,2$, see Fig.1:%
\begin{equation}
\mathbf{r}_{\mathbf{m}}^{1}=a\left( m_{1},\frac{1}{2}+m_{2}\right) ;\text{ \ 
}\mathbf{r}_{\mathbf{m}}^{2}=a\left( \frac{1}{2}+m_{1},m_{2}\right) \text{.}
\label{sublat}
\end{equation}%
Hopping occurs on each sublattice independently with amplitude $t$. The
overlap between nearest neighbors is negligible due to symmetry of orbitals.
In momentum representation on the 2D Brillouin zone (BZ), $-\pi
/a<k_{x},k_{y}<\pi /a$, one has (neglecting the spin $\sigma $ dependence%
\cite{DFT13}):

\begin{equation}
H_{e}=\sum \nolimits_{\mathbf{k}}c_{\mathbf{k}}^{A\sigma \dagger }\left(
\epsilon _{\mathbf{k}}-\epsilon _{F}\right) c_{\mathbf{k}}^{A\sigma }\text{,}
\label{tight1}
\end{equation}%
where $\epsilon _{\mathbf{k}}=-2t\left( 2+\cos \left[ ak_{x}\right] +\cos %
\left[ ak_{y}\right] \right) $. It is sufficient for our purposes to use a
parabolic approximation with an effective mass $m^{\ast }=2.7m_{e}$ and
Fermi energy\cite{Lee14} $\epsilon _{F}=60meV$. The electron gas is
considered non - interacting although screened delta - like repulsive
interaction should be added to the gap equation as discussed in\cite{our}.

\subsection{Optical phonon modes in the $TiO_{2}$\ layer.}

Phonons in ionic crystals are described by the Born - Meyer potential due to
electron's shells repulsion\cite{Abrahamson} and electrostatic interaction
of ionic charge,%
\begin{equation}
V^{XY}\left( r\right) =\sqrt{A_{X}A_{Y}}\exp \left[ \frac{1}{2}\left(
b^{X}+b^{Y}\right) r\right] +Z_{X}Z_{Y}\frac{e^{2}}{r}\text{,}
\label{interatomic}
\end{equation}%
with values of coefficients $A$ and $b$ listed in Table I. The ionic charges
of the STO plane below the last $TiO_{2}$ are taken from a DFT calculation 
\cite{averestov} of the Millikan charges (performed without $FeSe$). In the $%
TiO_{2}$ layer the charges are determined by neutrality, and a requirement
that the position of the oxygen atoms between the two $Ti$ atoms is a
minimum of potential.

It is reasonable to expect that the modes most relevant for the electron -
phonon coupling across the interface are the vibrations of the atoms in the $%
TiO_{2}$ layer, see Fig.5. Since oxygen is much lighter than $Ti$, we assume
that $Ti$ atoms' vibrations are negligible. Obviously we lose acoustic
branch, however the acoustic phonons are not expected to contribute to
pairing\cite{Mahan,Gorkov}. Atoms in neighboring layers can also be treated
as static. Moreover one can neglect more distant layers both in STO (beyond $%
SrO$) and in $FeSe$. Even the influence of the lower $Se$ layer is
insignificant due to the distance. Therefore the dominant lateral
displacements, $u_{\mathbf{m}}^{A\alpha }$, $\alpha =x,y$, are of the two
oxygen sublattices directly beneath the corresponding $Fe$ sites of Eq.(\ref%
{sublat}\textbf{)}. The dynamic matrix $\left[ D_{\mathbf{q}}\right]
_{\alpha \beta }^{AB}$ is calculated by expansion of energy to second order
in oxygen displacement (details in Appendix I), so that Hamiltonian is:

\begin{equation}
H_{ph}=\frac{1}{2}\sum \nolimits_{\mathbf{q}}\left \{ M\frac{du_{-\mathbf{q}%
}^{\alpha A}}{dt}\frac{du_{\mathbf{q}}^{\alpha A}}{dt}+u_{\mathbf{-q}%
}^{\alpha A}\left[ D_{\mathbf{q}}\right] _{\alpha \beta }^{AB}u_{\mathbf{q}%
}^{\beta B}\right \} \text{.}  \label{Hph}
\end{equation}%
Here $M$ is the oxygen mass. Summations over repeated sublattice and
components indices is implied. Now we turn to derivation of the phonon
spectrum and the electron - phonon coupling.

\section{Phonon spectrum and the electron - phonon interactions}

\subsection{Phonon spectrum}

Four eigenvalues are given in Fig. 6, while their polarization for a small $%
\mathbf{q}$ vector in $\mathbf{x}$ direction depicted in Fig. 5. One
observes that there are high and low frequency modes are in the range $%
\Omega _{\mathbf{q}}^{h}\sim 100-120mev$ and $\Omega _{\mathbf{q}}^{s}\sim $ 
$20-50mev$ respectively. The energy of LO modes (blue in Fig. 6) is larger
than that of the corresponding TO (red), although the sum $\Omega _{\mathbf{q%
}}^{LO}+\Omega _{\mathbf{q}}^{TO}$ is nearly dispersionless. At $\Gamma $
the splitting is small, while due to the long range Coulomb interaction
there is hardening of LO and softening of TO at the BZ edges. The dispersion
of the high frequency modes is small, while for the lower frequency mode it
is more pronounced.

\begin{figure}[h]
\centering \includegraphics[width=18cm]{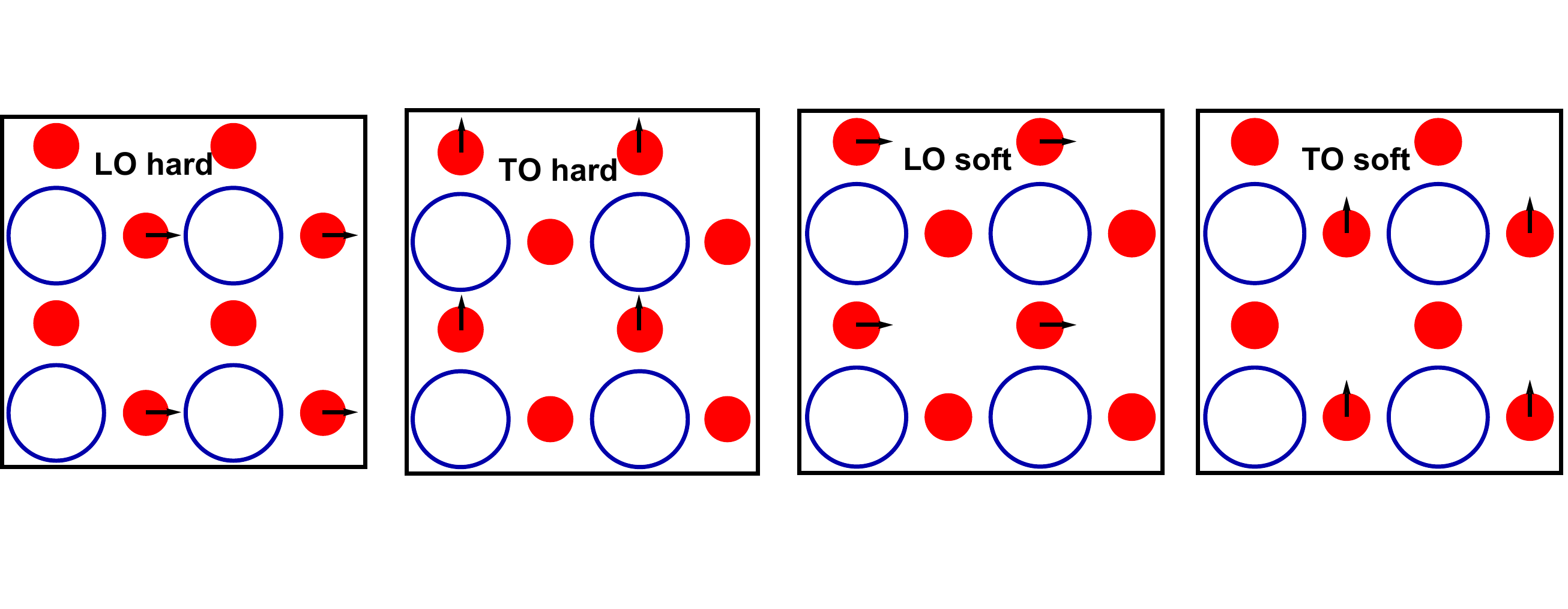}
\caption{ Oxygen atoms vibrations in the $TiO_{2}$ plane. Polarization of
the phonons with momentum along the $x$ axis. The oxygen atoms
displacement;s directions are indicated by arrows. Sublattice A=1 (see
Eq.(2)) is active in the hard TO and solf LO }
\end{figure}

\begin{figure}[h]
\centering \includegraphics[width=10cm]{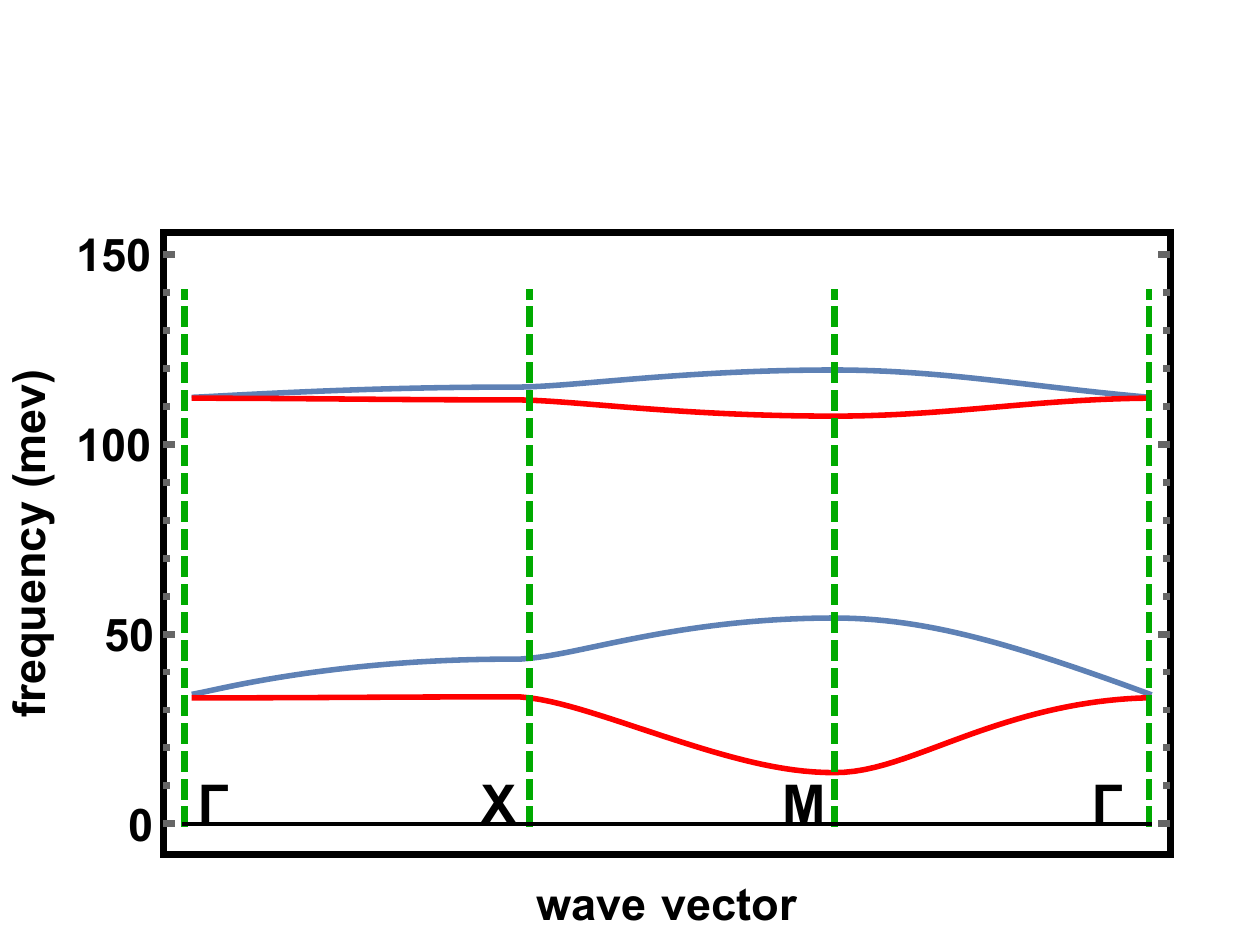}
\caption{ Spectrum of phonons in the $TiO_{2}$ plane. Blue lines correspond
to tranverse, while red line to the transverse modes.}
\end{figure}

Geometrically it is clear that the low frequency $\Omega _{\mathbf{q}}^{s}$
arises due the "empty site" at point $\left( 1/2,1/2\right) a$, Fig.1.
Physically the softness of the TO mode means that the crystal is close to
the ferroelectric transition of the displacement type characteristic to
oxides in perovskites\cite{ferroelectric} (the lowest frequency at the $M$
point of soft TO mode, $\Omega _{\mathbf{q=M}}^{sTO}$ would have reached
zero if the transition have occurred). Although the soft LO mode, $\Omega _{%
\mathbf{q}}^{sLO}$ (important for the electron - phonon coupling) is
slightly higher than $\Omega _{\mathbf{q}}^{sTO}$, it is still lower than $%
\Omega _{\mathbf{q}}^{h}$ $\ $by\ a significant factor $2.5$.

\subsection{Electron - phonon coupling}

The STO surface phonons interaction with the 2DEG on the $Fe$ layer $%
z_{Fe}=4.4A$ above the $TiO_{2}$ plane is determined by the electric
potential created near the $4d$ $Fe$ orbitals:

\begin{equation}
\Phi \left( r\right) =-Z_{O}e\sum \limits_{\mathbf{m,}A}\frac{\left( \mathbf{%
r}-\mathbf{r}_{\mathbf{m}}^{A}\right) \cdot \mathbf{u}_{\mathbf{m}}^{A}}{%
\left( \left( \mathbf{r}-\mathbf{r}_{\mathbf{m}}^{A}\right)
^{2}+z_{Fe}^{2}\right) ^{3/2}}\text{.}  \label{Potential}
\end{equation}%
It is important that the by vibrating charged oxygen atoms in the last $%
TiO_{2}$ layer reside directly below $Fe$ atoms. Influence on the electron -
phonon coupling of vibrating $Ti$ atoms of the first layer is further
reduced since they are not situated directly beneath the $Fe$ sites.

The potential generated by the charged $TiO_{2}$ oxygen vibration mode $u_{%
\mathbf{m}}^{A}$ at arbitrary point $\mathbf{r}$ is (namely ignoring
contributions from other charged ions is,

\begin{equation}
\Phi \left( \mathbf{r},z\right) =\sum \nolimits_{\mathbf{m,}A}\frac{Z_{O}%
\text{ }e}{\sqrt{\left( \mathbf{r}-\mathbf{r}_{m}^{A}-\mathbf{u}_{\mathbf{m}%
}^{A}\right) ^{2}+z^{2}}}\text{,}  \label{3_pot}
\end{equation}%
where the distance is to the $Fe$ layer, $z=z_{Fe}=4.4A.$ Expanding in
displacement, one \ obtains:

\begin{equation}
\Phi \left( \mathbf{r},z\right) \approx \Phi _{e-i}^{0}\left( \mathbf{r}%
,z\right) -Z_{O}\text{ }e\sum \nolimits_{\mathbf{m,}A}\frac{\left( \mathbf{r}%
-\mathbf{r}_{\mathbf{m}}^{A}\right) \cdot \mathbf{u}_{\mathbf{m}}^{A}}{%
\left( \left \vert \mathbf{r}-\mathbf{r}_{\mathbf{m}}^{A}\right \vert
^{2}+z^{2}\right) ^{3/2}}\text{.}  \label{3_potexpanded}
\end{equation}%
The Hamiltonian for interaction with electrons on the $4d$ $Fe$ orbitals
with wave functions $\varphi _{\mathbf{l}}^{A}\left( \mathbf{r},z\right) $
(on both sublattices $A=1,2$), $H_{ei}=\int_{\mathbf{r}}\Phi \left( \mathbf{r%
}\right) \widehat{n}_{\mathbf{r}}$, expanded to first order in the oxygen
vibrations consequently is,

\begin{equation}
H_{ei}=-Z_{O}e^{2}\int_{\mathbf{r},z}\sum \nolimits_{\mathbf{m,}A,B}\frac{%
\left( \mathbf{r}-\mathbf{r}_{\mathbf{m}}^{A}\right) \cdot \mathbf{u}_{%
\mathbf{m}}^{A}}{\left( \left( \mathbf{r}-\mathbf{r}_{\mathbf{m}}^{A}\right)
^{2}+z^{2}\right) ^{3/2}}\left \vert \varphi _{\mathbf{l}}^{A}\left( \mathbf{%
r},z\right) \right \vert ^{2}\widehat{c}_{\mathbf{l}}^{\sigma B\dagger }%
\widehat{c}_{\mathbf{l}}^{\sigma B}\text{.}  \label{3_Hei}
\end{equation}%
Sublattice indices are $A=1,2$ for oxygen and $B=1,2$ for $Fe$. Although the
most general matrix element depends also on the electron momentum $\mathbf{k}
$ in addition to the phonon momentum $\mathbf{q}$, it does not appear in Eq.%
\ref{3_Hei} since the coupling is to the density, namely the size of the $Fe$
orbital is neglected. Indeed the localized (the tight binding) form, namely,
neglecting the size of the orbital, $\left \vert \varphi _{\mathbf{l}%
}^{A}\left( \mathbf{r},z\right) \right \vert ^{2}=\delta \left( \mathbf{r-r}%
_{\mathbf{l}}^{B}\right) \delta \left( z-z_{Fe}\right) $, where $z_{Fe}$ is
given in Table I, reads:

\begin{equation}
H_{ei}=-Z_{O}e^{2}\int_{\mathbf{r},z}\sum \nolimits_{\mathbf{l},\mathbf{m,}%
A,B}\frac{\left( \mathbf{r}_{\mathbf{l}}^{B}-\mathbf{r}_{\mathbf{m}%
}^{A}\right) \cdot \mathbf{u}_{\mathbf{m}}^{A}}{\left( \left( \mathbf{r}-%
\mathbf{r}_{\mathbf{m}}^{A}\right) ^{2}+z_{Fe}^{2}\right) ^{3/2}}\widehat{n}%
_{\mathbf{l}}^{B\dagger }\text{.}  \label{3_localized}
\end{equation}%
Here the density operator $\widehat{n}_{\mathbf{l}}^{B\dagger }=\widehat{c}_{%
\mathbf{l}}^{\sigma B\dagger }\widehat{c}_{\mathbf{l}}^{\sigma B}$. The
interaction electron-phonon Hamiltonian has the form

\begin{equation}
H_{ei}=-e\int_{\mathbf{r}}\Phi \left( \mathbf{r}\right) \widehat{n}_{\mathbf{%
r}}=Z_{O}e^{2}\sum \nolimits_{\mathbf{q}}\widehat{n}_{-\mathbf{q}}^{B}g_{%
\mathbf{q}}^{BA\alpha }\widehat{u}_{\mathbf{q}}^{A\alpha }\text{,}
\label{Heph}
\end{equation}%
with $\widehat{n}_{\mathbf{q}}^{B}$ being Fourier transform of the electron
density operator on sublattice $B$ of $Fe$ and 
\begin{equation}
\mathbf{g}_{\mathbf{q}}^{BA}=\sum \nolimits_{\mathbf{m}}e^{ia\mathbf{q\cdot m%
}}\frac{\mathbf{r}_{\mathbf{m}}^{A}-\mathbf{r}_{\mathbf{0}}^{B}}{\left(
\left \vert \mathbf{r}_{\mathbf{m}}^{A}-\mathbf{r}_{\mathbf{0}}^{B}\right
\vert ^{2}+z_{Fe}^{2}\right) ^{3/2}}\text{,}  \label{A}
\end{equation}%
\cite{Grimvall,Eliashberg}. The later depends on sublattices of both the
vibrating oxygen atoms $A$ and the $Fe$ orbital hosting the electron on
sublattice $B$ (in addition to the polarization $\alpha $). It is well known
that only longitudinal phonons contribute to the effective electron -
electron interaction, as is clear from the scalar product form of the Eq.(%
\ref{Heph}). To conclude Eqs.(\ref{tight1},\ref{Hph},\ref{Heph}) define our
microscopic model. In order to describe superconductivity, one should
"integrate out" the phonon degrees of freedom to calculate the effective
electron - electron interaction. The discrete Fourier transform, 
\begin{equation}
\widehat{n}_{\mathbf{l}}^{B}=\frac{1}{N_{s}}\sum \nolimits_{\mathbf{q}}\exp %
\left[ -\frac{2\pi i}{N_{s}}\mathbf{q}\cdot l\right] \widehat{n}_{\mathbf{q}%
}^{B},  \label{3_Fourier}
\end{equation}%
together with Eq.(\ref{Heph}), result in the Matsubara action%
\begin{eqnarray}
\mathcal{A}_{eph}\left[ \psi ,u\right] &=&\frac{Z_{O}e^{2}}{T}\sum
\nolimits_{\mathbf{q,}n}n_{-\mathbf{q},-n}^{B}\left[ \psi \right] g_{\mathbf{%
q}}^{\alpha BA}u_{\mathbf{q},n}^{\alpha A}\text{;}  \label{3_acteph} \\
n_{-\mathbf{q},-n}^{B}\left[ \psi \right] &=&\sum \nolimits_{\mathbf{k,}%
m}\psi _{\mathbf{k}-\mathbf{q},m-n}^{\ast \sigma B}\psi _{\mathbf{k}%
,m}^{\sigma B}\text{.}  \notag
\end{eqnarray}%
that will be used below.

The electron - phonon coupling functions defined by,%
\begin{equation}
\rho _{\mathbf{q}}^{iA}\equiv \sum \nolimits_{C}\mathbf{e}_{\mathbf{q}%
}^{iC}\cdot \mathbf{g}_{\mathbf{q}}^{CA}\text{,}  \label{3_def}
\end{equation}%
depends on two indices, the phonon mode $i$ and a sublattice index $A$. The
"geometric" function $\mathbf{g}_{\mathbf{q}}^{CA}$ is defined in Eq.(8) of
the main text. 
\begin{figure}[h]
\centering \includegraphics[width=8cm]{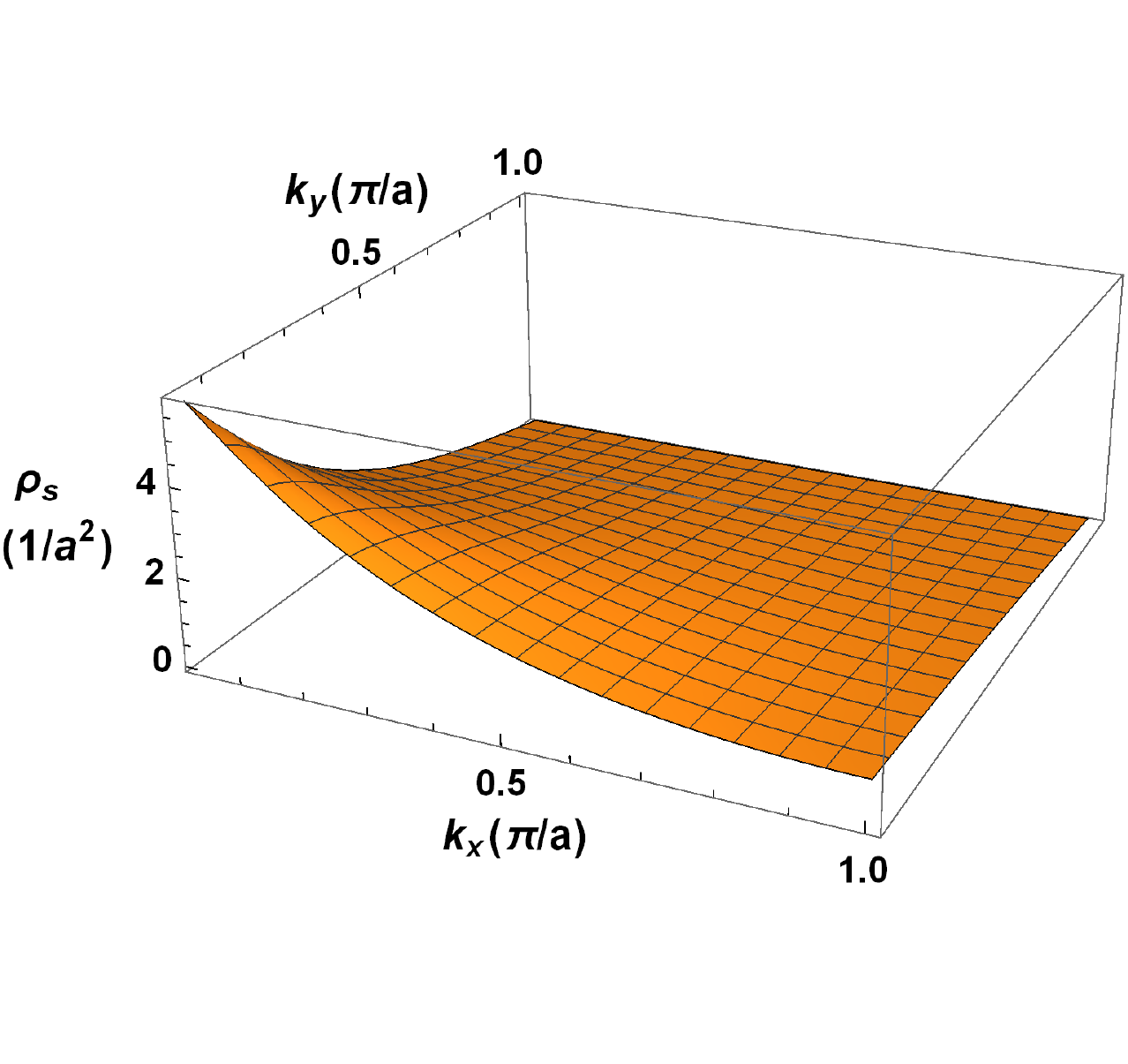}
\caption{ Electron - phonon coupling dependence on quasimomentum $\mathbf{q}$
(matrix element) on a quarter of Brillouin zone. The forward scattering peak
is clearly manifest.}
\end{figure}

\begin{figure}[h]
\centering \includegraphics[width=10cm]{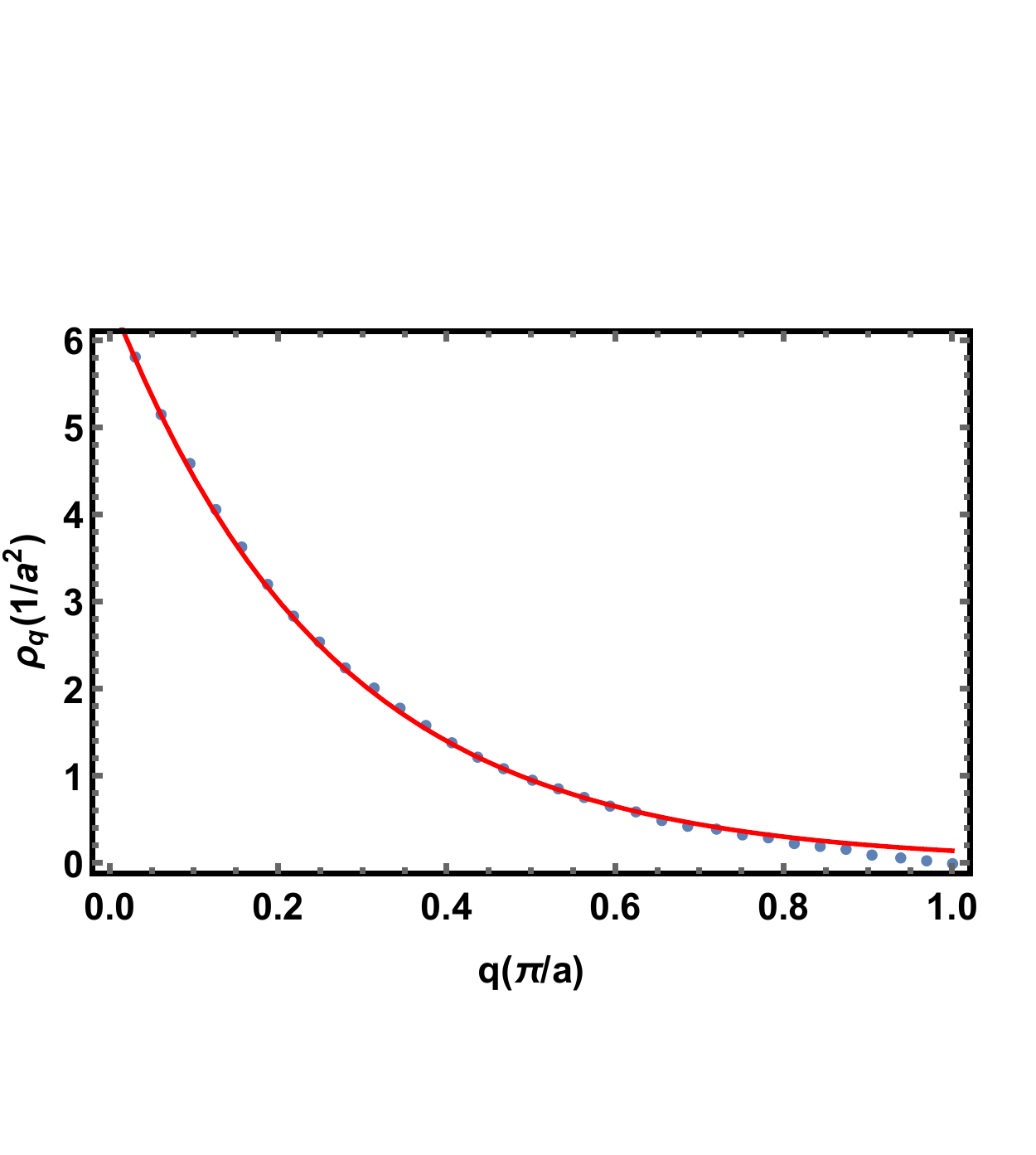}
\caption{ Fit of the electron - phonon nteraction strangth of the soft mode
by an exponential function of Eq. 18of the main text. }
\end{figure}
The corresponding plots for sublattice $A=2$ are rotated by $\pi /2$ due to
the fourfold symmetry. The continuos rotation symmetry is weakly broken at
edges of the Brillouin zone. The shape is slightly different for the hard
and soft mode, however the rotation invariant fit of Eq.(16) is correct to $%
5\%$ as seen in Fig. 7. The transversal modes are smaller by an order of
magnitude.

The mostly transversal contributions $\rho _{\mathbf{q,}h}^{A}$ and $\rho _{%
\mathbf{q,}s}^{A}$ are negligible (albeit nonzero for general $\mathbf{q}$
due to lack of continuous rotational symmetry). The LO contributions can be
approximated within 1\% (see Fig.8) by%
\begin{equation}
\rho _{\mathbf{q}}^{A}\approx \rho e^{-\left \vert \mathbf{q}\right \vert
/q_{0}}\text{,}  \label{rho}
\end{equation}%
with $\rho =2\pi /a^{2}\ $and $q_{0}=1/z_{Fe}=0.9/a\approx 0.23A^{-1}$ for
both modes. The exponential decrease reflects\cite{Lee12,JohnsonNJP16,Kulic}
the distance between the phonon layer and the 2DEG.

\subsection{Effective electron - electron interaction\textit{. }}

\textit{\ }To take into account finite temperature, we employ the Matsubara
action\cite{our} for the above Hamiltonian, $\mathcal{A}=\mathcal{A}_{e}+%
\mathcal{A}_{ph}+\mathcal{A}_{eph}$, where 
\begin{eqnarray}
\mathcal{A}_{e} &=&T^{-1}\sum \nolimits_{\mathbf{k},n}\psi _{\mathbf{k}%
,n}^{\ast \sigma A}\left( G_{\mathbf{k},n}^{0}\right) ^{-1}\psi _{\mathbf{k}%
,n}^{\sigma A}\text{, }\mathcal{A}_{ph}=\frac{M}{2T}\sum \nolimits_{\mathbf{%
q,}n}u_{-\mathbf{q},-n}^{\alpha A}\left[ \Pi _{\mathbf{q},n}\right] _{\alpha
\beta }^{AB}u_{\mathbf{q,}n}^{\beta B};  \label{action} \\
\text{ }\mathcal{A}_{eph} &=&\frac{Z_{O}e^{2}}{T}\sum \nolimits_{\mathbf{q,}%
n}n_{-\mathbf{q},-n}^{B}g_{\mathbf{q}}^{\alpha BA}u_{\mathbf{q},n}^{\alpha A}%
\text{.}  \notag
\end{eqnarray}%
Here the bare Green's function for normal electrons described by a
Grassmanian field $\psi $, is,%
\begin{equation}
G_{\mathbf{k},n}^{0}=\left( i\omega _{n}^{f}-\epsilon _{\mathbf{k}}+\epsilon
_{F}\right) ^{-1}\text{,}  \label{G0}
\end{equation}%
with $\omega _{n}^{f}=\pi T\left( 2n+1\right) $. Here the density is written
in terms of. The $4\times 4$ polarization matrix, 
\begin{equation}
\left[ \Pi _{\mathbf{q},n}\right] _{\alpha \beta }^{AB}=\left( \omega
_{n}^{b}\right) ^{2}\delta ^{AB}\delta _{\alpha \beta }+M^{-1}\left[ D_{%
\mathbf{q}}\right] _{\alpha \beta }^{AB}\text{,}  \label{Pi}
\end{equation}%
is defined via the dynamic matrix of Eq.(\ref{Hph}) calculated in Appendix I
and $\omega _{n}^{b}=2\pi nT$ is the Matsubara frequency for phonons. The
action is completed by the free electron action,

Since the action is quadratic in the phonon field $\mathbf{u}$ the partition
function is gaussian, it can be integrated out exactly. The electronic
effective action is obtained by integration of the partition function over
the phonon field,

\begin{equation}
e^{-\mathcal{A}_{eff}\left[ \psi \right] }=\int_{u}e^{-\mathcal{A}_{ph}\left[
u\right] -\mathcal{A}_{eph}\left[ \psi ,u\right] }\text{,}  \label{3_action}
\end{equation}%
where the phonon action is 
\begin{equation}
\mathcal{A}_{ph}=\frac{M}{2T}\sum \nolimits_{\mathbf{q,}n}u_{-\mathbf{q}%
,-n}^{\alpha A}\left[ \Pi _{\mathbf{q},n}\right] _{\alpha \beta }^{AB}u_{%
\mathbf{q,}n}^{\beta B}\text{,}  \label{3_actph}
\end{equation}%
and the electron - phonon part is given by Eq.(10) .

The integral is gaussian in the fields $u_{\mathbf{q}n}^{\beta B}$ and thus,
since normalization constant is independent of the electron field in $n_{-%
\mathbf{q}n}^{B}$, is performed by completion to full square. The result
collecting the constants is,%
\begin{equation}
e^{-\mathcal{A}_{eff}\left[ \psi \right] }\propto \exp \left[ -\frac{\left(
Z^{O}e^{2}\right) ^{2}}{2MT}\sum \nolimits_{\mathbf{q,}n}n_{\mathbf{q}%
,n}^{B}g_{\mathbf{q}}^{BC\gamma }\left[ \Pi _{\mathbf{q},n}^{-1}\right]
_{\gamma \delta }^{CD}g_{-\mathbf{q}}^{DA\delta }n_{-\mathbf{q},-n}^{A}%
\right] \text{.}  \label{4_completion}
\end{equation}%
As a result one obtains the effective density - density interaction term for
of electrons

\begin{equation}
\mathcal{A}_{eff}=\frac{1}{2T}\sum \nolimits_{\mathbf{q}.n}n_{\mathbf{q}%
,n}^{B}v_{\mathbf{q},n}^{BA}n_{-\mathbf{q},-n}^{A}\text{,}  \label{Aee}
\end{equation}%
where the effective electron - electron frequency dependent potential is%
\begin{equation}
v_{\mathbf{q},n}^{BA}=-\frac{\left( Z_{O}e^{2}\right) ^{2}}{M}g_{\mathbf{q}%
}^{BC\gamma }\left[ \Pi _{\mathbf{q},n}^{-1}\right] _{\gamma \delta
}^{CD}g_{-\mathbf{q}}^{DA\delta }\text{.}  \label{Aeff}
\end{equation}%
In the basis of the four phonon modes with polarization vectors $\mathbf{e}_{%
\mathbf{q}}^{iC}$ depending on the phonon branch $i=1,...4$, this becomes:%
\begin{equation}
v_{\mathbf{q},n}^{BA}=-\frac{\left( Z_{O}e^{2}\right) ^{2}}{M}\sum
\nolimits_{i=1}^{4}\frac{\rho _{\mathbf{q,}i}^{B}\rho _{-\mathbf{q,}i}^{A}}{%
\omega _{n}^{b2}+\Omega _{\mathbf{q,}i}^{2}}\text{; \ }\rho _{\mathbf{q,}%
i}^{A}\equiv \sum \nolimits_{C}\mathbf{e}_{\mathbf{q}}^{iC}\cdot \mathbf{g}_{%
\mathbf{q}}^{CA}\text{.}  \label{vfinal1}
\end{equation}

Consequently Eq.(\ref{vfinal1}) takes a form:%
\begin{equation}
v_{\mathbf{q},n}^{AB}\approx -\frac{\left( Z_{O}e^{2}\rho \right) ^{2}}{M}%
e^{-2\left \vert \mathbf{q}\right \vert /q_{0}}\left( \frac{1}{\omega
_{n}^{b2}+\Omega _{\mathbf{q,}h}^{2}}+\frac{1}{\omega _{n}^{b2}+\Omega _{%
\mathbf{q,}s}^{2}}\right) \text{,}  \label{vdominant}
\end{equation}%
approximately independent of sublattice indices. One observes that at $%
nT<<\Omega $ the dominant mode one is the soft LO mode for superconductivity
and even for satellites.

\section{Superconductivity.\textit{\ }}

\subsection{Gap equation}

The STM experiments\cite{swave} demonstrate that the order parameter is
gapped (hence no nodes) and indicate a weakly anisotropic spin singlet
pairing. Therefore we look for solutions for the normal and the anomalous
Green's function of the Gorkov equations in the form

\begin{equation}
\left \langle \psi _{\mathbf{k},n}^{\rho B}\psi _{\mathbf{k},n}^{\ast \sigma
A}\right \rangle =\delta ^{\sigma \rho }G_{\mathbf{k},n}^{AB};\text{ }\left
\langle \psi _{\mathbf{k},n}^{\sigma A}\psi _{-\mathbf{k},-n}^{\rho B}\right
\rangle =\varepsilon ^{\sigma \rho }F_{\mathbf{k},n}^{AB}  \label{GFdef1}
\end{equation}%
where $\varepsilon ^{\sigma \rho }$ is the antisymmetric tensor. At
criticality, $G_{\mathbf{k},n}^{AC}=\delta ^{AC}G_{\mathbf{k},n}^{0}$
(normal Green's function not renormalized significantly at weak coupling),
the Gorkov equation for the anomalous Greens function is (derived for a
multi - band system in Appendix B):%
\begin{equation}
F_{\mathbf{p},m}^{AC}+T_{c}\left \vert G_{\mathbf{p},m}^{0}\right \vert
^{2}\sum \nolimits_{\mathbf{q},n}v_{\mathbf{p-q},m-n}^{AC}F_{\mathbf{q}%
,n}^{AC}=0\text{.}  \label{secondG}
\end{equation}%
In terms of the gap function,

\begin{equation}
\Delta _{\mathbf{p},m}^{AC}=T_{c}\sum \nolimits_{\mathbf{k},n}v_{\mathbf{p-k}%
,m-n}^{AC}F_{\mathbf{k},n}^{AC}\text{,}  \label{deltadef}
\end{equation}%
this becomes

\begin{equation}
T_{c}\sum \nolimits_{\mathbf{p},m}\left \vert g_{\mathbf{p},m}^{0}\right
\vert ^{2}v_{\mathbf{q-p},n-m}^{AC}\Delta _{\mathbf{p},m}^{AC}=-T_{c}\sum%
\nolimits_{\mathbf{p},m}\frac{v_{\mathbf{q-p},n-m}^{AC}}{\left( \omega
_{m}^{e}\right) ^{2}+\left( \epsilon _{\left \vert \mathbf{p}\right \vert
}-\epsilon _{F}\right) ^{2}}\Delta _{\mathbf{p,}m}^{AC}=\Delta _{\mathbf{q}%
,n}^{AC}\text{.}  \label{gapeq}
\end{equation}

From this point on let us assume that we consider just the dominant mode and
that this mode is dispersionless., see Eq.(\ref{vdominant}) $\Omega _{%
\mathbf{q}s}=\Omega _{s}$. In addition only element $v_{\mathbf{q}%
,n}^{11}=v_{\mathbf{q},n}^{22}$ is considered, so that the sublattice index
will be omitted. The resulting sum near a circular Fermi surface can be
approximated by an integral:%
\begin{equation}
\frac{T_{c}\left( Z_{O}e^{2}\rho \right) ^{2}}{M}\sum \nolimits_{m}\frac{1}{%
\omega _{n-m}^{b2}+\Omega ^{2}}\frac{1}{\left( 2\pi \right) ^{2}}\int_{%
\mathbf{p}}\frac{e^{-2\left \vert \mathbf{q-p}\right \vert /q_{0}}}{\left(
\omega _{m}^{e}\right) ^{2}+\left( \epsilon _{\left \vert \mathbf{p}\right
\vert }-\epsilon _{F}\right) ^{2}}\Delta _{\mathbf{p,}m}=\Delta _{\mathbf{q,}%
n}\text{.}  \label{4_gapeq}
\end{equation}%
Using rotation invariance one obtains the following gap equation for an
angle independent gap function, $\Delta _{\mathbf{p,}m}=\Delta _{p\mathbf{,}%
m}$, in polar coordinates:%
\begin{equation}
\frac{T_{c}\left( Z_{O}e^{2}\rho \right) ^{2}}{M\left( 2\pi \right) ^{2}}%
\sum \nolimits_{m}\frac{1}{\omega _{n-m}^{b2}+\Omega ^{2}}\int_{p=0}\frac{p%
\text{ \ }\gamma \left[ p,q\right] }{\left( \omega _{m}^{e}\right)
^{2}+\left( \epsilon _{p}-\epsilon _{F}\right) ^{2}}\Delta _{p\mathbf{,}%
m}=\Delta _{q\mathbf{,}n}\text{.}  \label{gapexact}
\end{equation}%
The integration over the difference of angles can be performed numerically,%
\begin{equation}
\gamma \left[ p,q\right] =\int_{\phi =0}^{2\pi }\exp \left[ -2\sqrt{%
p^{2}+q^{2}-2pq\cos \phi }/q_{0}\right] \text{.}  \label{angle}
\end{equation}%
This eigenvalue problem was first solved numerically and then (in Appendix
C) within the Eliashberg approximation in the case when the main
contribution comes from momenta very close to $k_{F}$. Both methods gives
the same value for the critical temperature $T_{c}=51K$

\subsection{Solution of the gap equation}

Momenta within the circular Brillouin zone of radius $\pi /a$ were
discretized as $p\rightarrow \frac{\pi }{N_{s}a}p$\ with $N_{s}=400$, while
the Matsubara frequency $m$ was truncated at $\left \vert \pi T_{c}\left(
2m-1\right) \right \vert \simeq 4\Omega _{s}$. Time reversal symmetry
ensures $\Delta _{p,-m}=\Delta _{p,m+1}$, so that only positive integers
were included%
\begin{eqnarray}
K_{pm,qn}\Delta _{p\mathbf{,}m} &=&\Delta _{q\mathbf{,}n}  \label{matrixK} \\
K_{pm,qn} &=&\frac{T_{c}\left( Z_{O}e^{2}\rho \right) ^{2}}{MN_{s}}\frac{p%
\text{ \ }\gamma \left[ p,q\right] }{\left( \omega _{m}^{e}\right)
^{2}+\left( \epsilon _{p}-\epsilon _{F}\right) ^{2}}  \notag
\end{eqnarray}%
The critical temperature is obtained when the largest eigenvalue of the
matrix $K$ Eq.( \ref{gapexact}) is unit. The numerical results are the
following. $T_{c}=51K$, while for isotope $^{18}O$ it becomes $T_{c}=49K$.

It is clearly demonstrated in Fig. 9 that the dependence on $m$ is very
strong: the two lowest Matsubara frequencies $\Delta _{m}$, $m=0,1$ for
which $\left \vert \pi T\left( 2m-1\right) \right \vert =\pi T$ are
dominant, while corrections of $m=-3,4$ (yellow line in Fig. 9)\ become less
than 1\%.

\begin{figure}[h]
\centering \includegraphics[width=16cm]{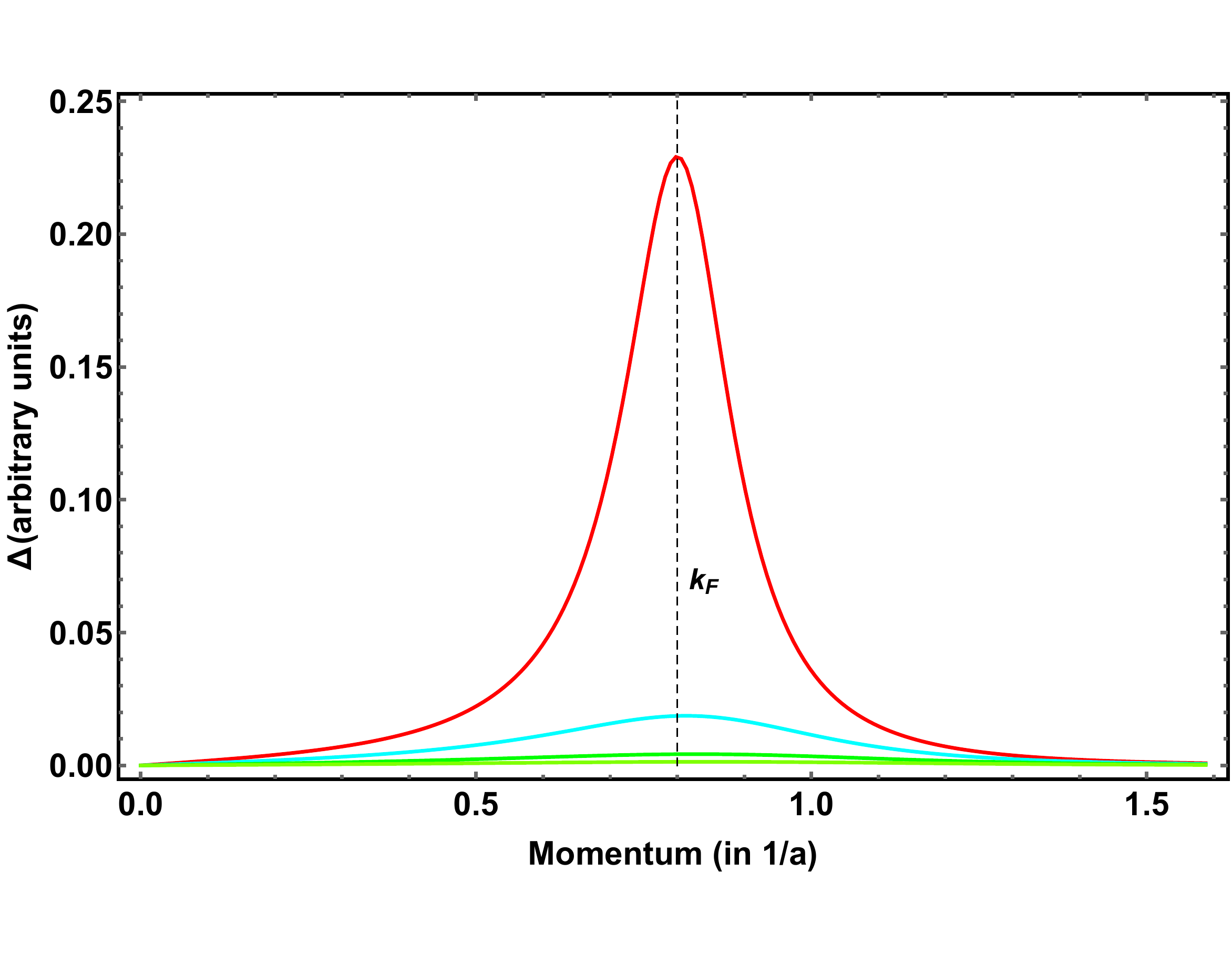}
\caption{ Gap function for different Matsubara frequences The m=0,1 (red) is
dominant, while strength of the subleading correlators, m=-1,2 (cyan),
m=-2,3 (green), m=-3,4 (yellow), decrease fast.}
\end{figure}

Shape of the momenta distribution of all the modes can be described as a
Lorenzian around $k_{F}$. The width is significant due to exceptionally
small "adiabaticity parameter" $E_{F}/\Omega _{s}=1.4$ (would be smaller for
the hard mode $\Omega _{h}$). The Lorenzian width shrinks to zero for small $%
q_{0}$ (the delta forward peak scattering limit) and for adiabatic case of
large $E_{F}/\Omega _{s}$. The gap function $\Delta _{k,\omega }$ vanishes
at the transition temperature and increases below it as $\left(
T-T_{c}\right) ^{1/2}$ according to Ginzburg - Landau approach preserving
its shape.

The dominant region around $k_{F}$ allows application of the Eliashberg
approximation, that in the present case allows analytic solution presented
in Appendix C. The results are consistent with numerical simulation.

\section{Normal state effects of the electron - phonon interactions.}

\subsection{Self Energy}

The first Gorkov equation in the normal phase, namely Eq.(\ref{4_Gor1}) for
anomalous Green function $F=0$, is just the conventional gaussian
approximation:

\begin{equation}
G_{\alpha }^{CA}+G_{\alpha }^{0}G_{\alpha }^{BA}v_{\chi }^{BC}G_{\chi
+\alpha }^{BC}-2G_{\alpha }^{0}G_{\alpha }^{CA}v_{0}^{CX}G_{\chi
}^{XX}=\delta ^{AC}G_{\alpha }^{0}\text{.}  \label{5_gauss}
\end{equation}%
The sublattice Ansatz, $G^{AB}=\delta ^{AB}G$, already used at the critical
point is still valid,

\begin{equation}
\delta ^{AC}\left( G_{\alpha }+G_{\alpha }^{0}G_{\alpha }v_{\chi
}^{AA}G_{\chi +\alpha }-2G_{\alpha }^{0}G_{\alpha }G_{\chi }\sum
\nolimits_{X}v_{0}^{AX}\right) =\delta ^{AC}G_{\alpha }^{0}\text{,}
\label{5_Ansatz}
\end{equation}%
since $v^{11}=v^{22},v^{12}=v^{21}$ due to the fourfold symmetry.
Consequently in components one can write:%
\begin{equation}
G_{\alpha }^{-1}=\left( G_{\alpha }^{0}\right) ^{-1}+\sum \nolimits_{\chi
}v_{\chi }^{11}G_{\chi +\alpha }-2\left( v_{0}^{11}+v_{0}^{12}\right) \sum
\nolimits_{\chi }G_{\chi }\text{.}  \label{5_sedef}
\end{equation}

The\ frequency-momentum\ independent\ term\ is\ accounted\ for\ by\
renormalization\ of\ the\ chemical\ potential. While\ in\ principle\ this\
equation should\ be\ solved\ self\ consistently,\ since\ the\ electron\ -\
phonon\ interaction\ is\ relatively\ weak,\ one\ neglects\ the\ correction\
to\ $G_{\chi +\alpha }^{0}$ on\ the\ right\ hand\ side.\ This\ results\ in\
the\ perturbation\ theory\ formula for the self\ energy (substituting\ the\
expression\ for\ $v_{\chi }^{11}$\ from\ Eq.$\left( \text{\ref{rho}}\right) $
and $G^{0}$ from Eq.(\ref{G0})\ :

\begin{equation}
\Sigma \left( n,\mathbf{k}\right) =-\sum \nolimits_{\chi }v_{\chi
}^{11}G_{\chi +\alpha }^{0}=\frac{\left( Z^{O}e^{2}\right) ^{2}T}{MN_{s}^{2}}%
\sum \nolimits_{\mathbf{l,}m,i}\frac{\left \vert \rho _{\mathbf{l,}%
i}^{1}\right \vert ^{2}}{\left( \left( 2\pi Tm\right) ^{2}+\Omega _{\mathbf{l%
}}^{i2}\right) \left( i\pi T\left( 2m+2n+1\right) -\left( \varepsilon _{%
\mathbf{k+l}}-\epsilon _{F}\right) \right) }  \label{4_sigmadef}
\end{equation}%
Summing\ over\ the bosonic\ Matsubara\ frequency\ $m$,\ one\ obtains:

\begin{equation}
\Sigma \left( n,\mathbf{k}\right) =\frac{\left( Z^{O}e^{2}\right) ^{2}}{%
MN_{s}^{2}}\sum \nolimits_{\mathbf{l,}i}\frac{\left \vert \rho _{\mathbf{l,}%
i}^{1}\right \vert ^{2}}{2\Omega _{\mathbf{l}}^{i}}\left \{ \frac{n_{b}\left[
\Omega _{\mathbf{l}}^{i}\right] -n_{f}\left[ -\left( \varepsilon _{\mathbf{%
k+l}}-\epsilon _{F}\right) \right] +1}{i\omega _{n}^{f}+\Omega _{\mathbf{l}%
}^{i}-\left( \varepsilon _{\mathbf{k+l}}-\epsilon _{F}\right) }+\frac{n_{b}%
\left[ \Omega _{\mathbf{l}}^{i}\right] +n_{f}\left[ -\left( \varepsilon _{%
\mathbf{k+l}}-\epsilon _{F}\right) \right] }{i\omega _{n}^{f}-\Omega _{%
\mathbf{l}}^{i}-\left( \varepsilon _{\mathbf{k+l}}-\epsilon _{F}\right) }%
\right \}  \label{4_sig}
\end{equation}%
where the Bose and the Fermi distributions are%
\begin{equation}
n_{b}\left[ e\right] =\frac{1}{\exp \left[ e/T\right] -1};\text{ \  \ }n_{f}%
\left[ e\right] =\frac{1}{1+\exp \left[ e/T\right] }.
\label{5_distributions}
\end{equation}%
This is used below to calculate the dimensionless coupling constant $\lambda 
$ and to describe the "satellites" in the electron spectrum.

For momentum on the Fermi surface, $\varepsilon _{\mathbf{k}}=\varepsilon
_{F}$ one can use the parabolic band approximation formula Eq.(\ref{3_ksidef}%
). At low temperatures (compared to $\Omega $) retaining a single mode with
frequency $\Omega $, the self energy (utilizing the interpolation formula of
Eq.(\ref{vdominant}) for $\rho _{\mathbf{l,}i}^{1}$) takes a form (replacing
the sum over momenta $\mathbf{l}$ by an integral in polar coordinates $%
l,\phi $),

\begin{equation}
\Sigma \left( \omega \right) =\frac{\left( Z_{O}e^{2}\rho \right) ^{2}}{8\pi
^{2}M\Omega }\int_{l=0}^{2k_{F}}le^{-2l/q_{0}}\int_{\phi =0}^{2\pi }\left( 
\frac{\Theta \left[ -\xi _{l,\phi }\right] }{i\omega +\Omega -\xi _{l,\phi }}%
+\frac{\Theta \left[ \xi _{l,\phi }\right] }{i\omega -\Omega -\xi _{l,\phi }}%
\right) \text{,}  \label{4_polar}
\end{equation}%
where $\Theta $ is the Heaviside function and $\xi _{l,\phi }$ was defined
as 
\begin{equation}
\xi _{l,\phi }\equiv \varepsilon _{\mathbf{q+l}}-\epsilon _{F}=\frac{%
l+2k_{F}\cos \phi }{2m^{\ast }}l\text{.}  \label{3_ksidef}
\end{equation}
As above one incorporates for example the step function $\Theta \left[ -\xi
_{l,\phi }\right] $ as a restriction on the integration range, $l+2k_{F}\cos
\phi <0$, leading the the limiting value of $\phi _{0}=\arccos \left[ -%
\widetilde{l}\right] $ with dimensionless momentum $\widetilde{l}\equiv
l/2k_{F}$.

Let us transform the self energy to physical (dimensionless) frequencies as $%
\frac{m^{\ast }}{2k_{F}^{2}}i\omega \rightarrow w+i\eta $ for infinitesimal
positive $\eta $. The self energy takes a form (the tilde over $l$ is
suppressed from now on):

\begin{equation}
\Sigma \left( w\right) =\frac{g^{2}}{\widetilde{\Omega }}\frac{2k_{F}^{2}}{%
m^{\ast }}\int_{l=0}^{1}e^{-4k_{F}l/q_{0}}\left \{ I_{1}\left[ l-\left(
w+i\eta +\widetilde{\Omega }\right) /l\right] +I_{2}\left[ l+\left( 
\widetilde{\Omega }-w-i\eta \right) /l\right] \right \} \text{,}
\label{4_scaled}
\end{equation}%
where $\widetilde{\Omega }=\frac{m^{\ast }}{2k_{F}^{2}}\Omega $, and the
electron - phonon coupling constant is defined as,%
\begin{equation}
g^{2}=\frac{\left( Z_{O}e^{2}\rho \right) ^{2}m^{\ast 3}}{8\pi ^{2}Mk_{F}^{4}%
}\text{.}  \label{4_dimensionless}
\end{equation}%
The angle integrals in Eq.(\ref{4_polar}) were performed for any complex
parameter $a$:

\begin{eqnarray}
I\left[ a\right] &=&-\int_{\phi =0}^{\pi }\frac{1}{a+\cos \phi }=-(-1)^{%
\text{int}\left[ \left( -2\text{arg}[-1+a]+\text{arg}[1-a^{2}]\right) /2\pi %
\right] }\frac{i\pi }{\sqrt{1-a^{2}}};  \label{4_Idefine} \\
I_{2}\left[ a\right] &=&-\int_{\phi =0}^{\arccos \left[ -l\right] }\frac{1}{%
a+\cos \phi }=\frac{2}{\sqrt{1-a^{2}}}\text{arctanh}\left[ \frac{a-1}{\sqrt{%
1-a^{2}}}\tan \left[ \frac{\phi _{0}}{2}\right] \right] ;  \notag \\
I_{1}\left[ a\right] &=&I\left[ a\right] -I_{2}\left[ a\right] \text{.} 
\notag
\end{eqnarray}%
This expression will be used for description of the ARPES satellites and the
effective electron - electron dimensionless coupling $\lambda $.

\subsection{Quasiparticle spectrum and satellites}

The spectral weight of quasiparticles (electrons) is given by the imaginary
part of the full Green function containing the effects of the electron -
phonon interaction

\begin{equation}
A_{\mathbf{k}}\left( w\right) =-\frac{1}{\pi }\text{Im }G\left( w,\mathbf{k}%
\right) .  \label{4_spec}
\end{equation}%
For momentum on the Fermi surface, $\varepsilon _{\mathbf{k}}=\varepsilon
_{F}$, using the self energy of Eq.(\ref{4_scaled}), it is%
\begin{equation}
G\left( w\right) =\frac{1}{w+i\eta -\Sigma \left( w\right) }\text{.}
\label{4_G}
\end{equation}

The spectral weight is presented in Fig. 5 for $\Omega =\Omega _{s}$ (from
now on we drop tilde, $\widetilde{\Omega }=\frac{m^{\ast }}{2k_{F}^{2}}%
\Omega \rightarrow \Omega $) and $\eta =0.03$. One observes that beyond the
dominant sharp quasiparticle peak near $w=0$, there are two small
"satellite" structures created by the soft phonon mode. The one observed on
ARPES extends from the phonon mode $w=-\Omega _{s}$ all the way to the peak
at the satellite location slightly above $w=-\Omega _{s}-1/4$.

The location of the "satellite" (poles) is determined by solving the
equation for diverging normal Greens function for physical frequencies. 
\begin{equation}
w=\frac{g^{2}}{\Omega }\int_{l=0}^{1}e^{-4k_{F}l/q_{0}}\text{Re}\left[ I_{1}%
\left[ l-\left( w-\Omega \right) /l\right] +I_{2}\left[ l-\left( w-\Omega
\right) /l\right] \right] \text{.}  \label{4_equation}
\end{equation}%
The small imaginary part $i\eta $ is not required since expressions in Eq.(%
\ref{4_Idefine}) reproduce exactly the principal value integrals over $l$ in
Eq.(\ref{4_scaled}). The integrand of the RHS of the equation (the self
energy), is an integrable discontinuous function. It is given in Fig. 10.

\begin{figure}[h]
\centering \includegraphics[width=10cm]{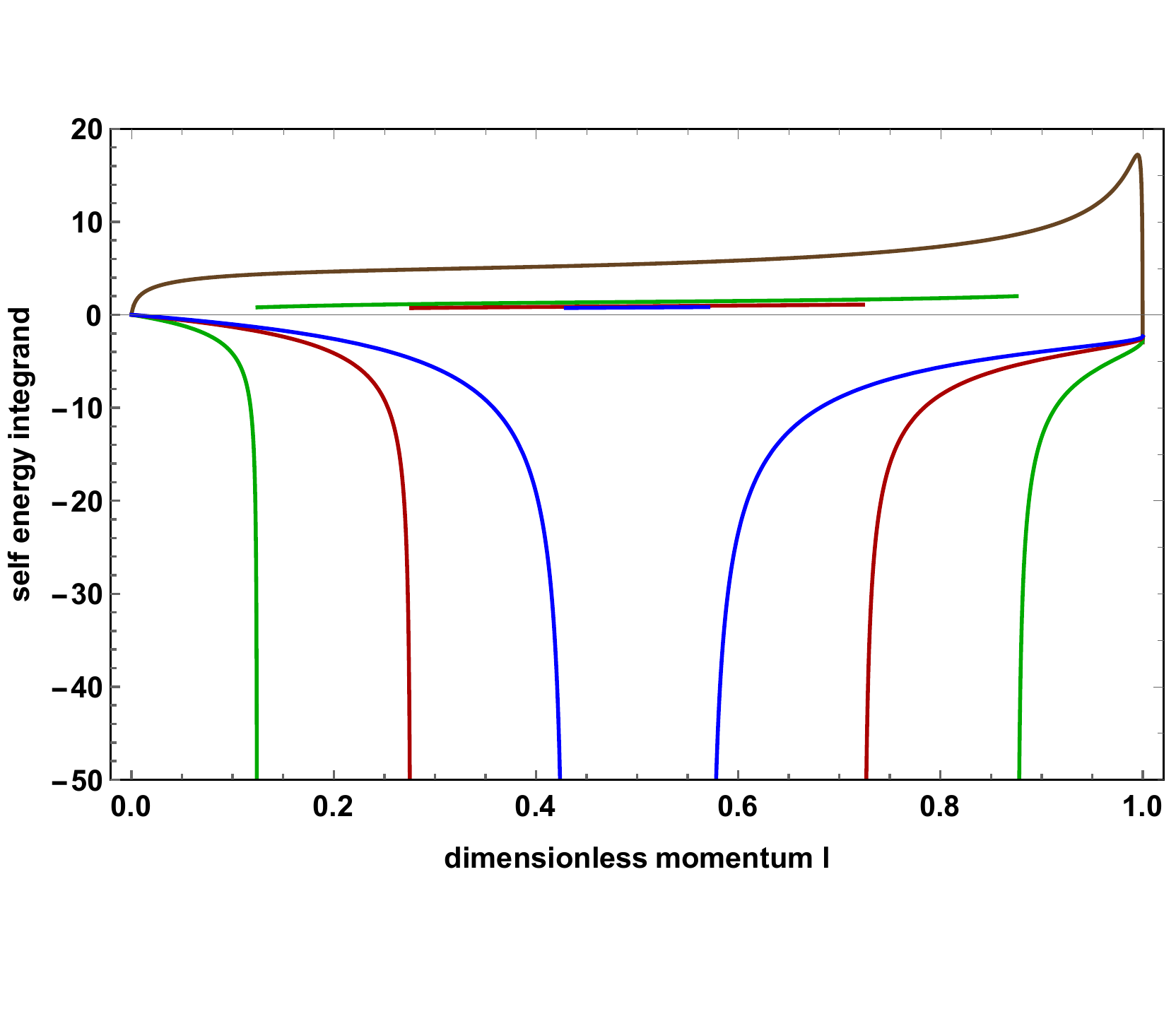}
\caption{Discontinuities of the integrand over quasimomentum $l$. \ The
jumps appear at two points, Eq.(\protect \ref{l12}) for all frequencies
between the location of the ARPES satellite peak and negative phonon
frequency $-\Omega _{s}$.}
\end{figure}

There are discontinuities at 
\begin{equation}
l_{1,2}=\frac{1}{2}\left( 1\pm \sqrt{1+4\left( w+\Omega \right) }\right) ,
\label{l12}
\end{equation}
when the argument of function $I_{1}\left( a\right) $ equals $1$. The
integration was performed in any region separately.

\begin{figure}[h]
\centering \includegraphics[width=10cm]{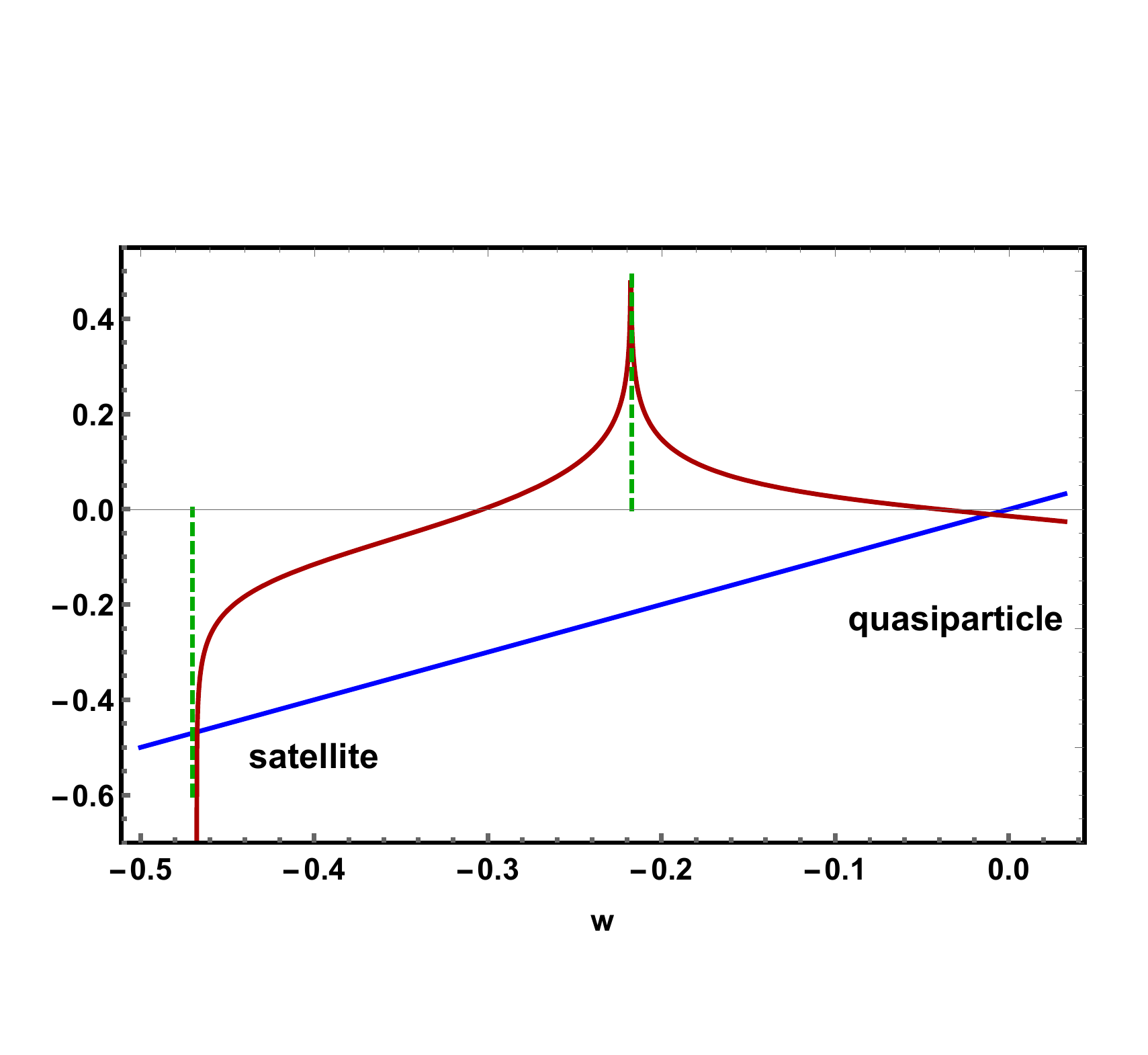}
\caption{Graphcal soluton of the $w=\Sigma \left( k_{F},w\right) $ equation.
Two solutions corresponding to the ARPES satellite peak (negative energy)
and the main quasiparticle excitation near zero are apparent.}
\end{figure}
It is important to note that the discontinuity disappears at $l=1/2$ when $%
1+4\left( w+\Omega \right) =0$,\ determining the discontinuity of the
integral to be at $w^{-}=-\Omega -1/4$ The equation is solved graphically in
Fig. 11 and numerically in Fig. 12 \ for $\Omega _{h}$ and $\Omega _{s}$.
Returning to physical units, for $\Omega _{s}=51.6mev$ one obtains $\omega
^{-}=-90.0mev$ with divergence of the spectral weight appearing at $\omega
^{peak}=-89.5mev$.

\begin{figure}[h]
\centering \includegraphics[width=16cm]{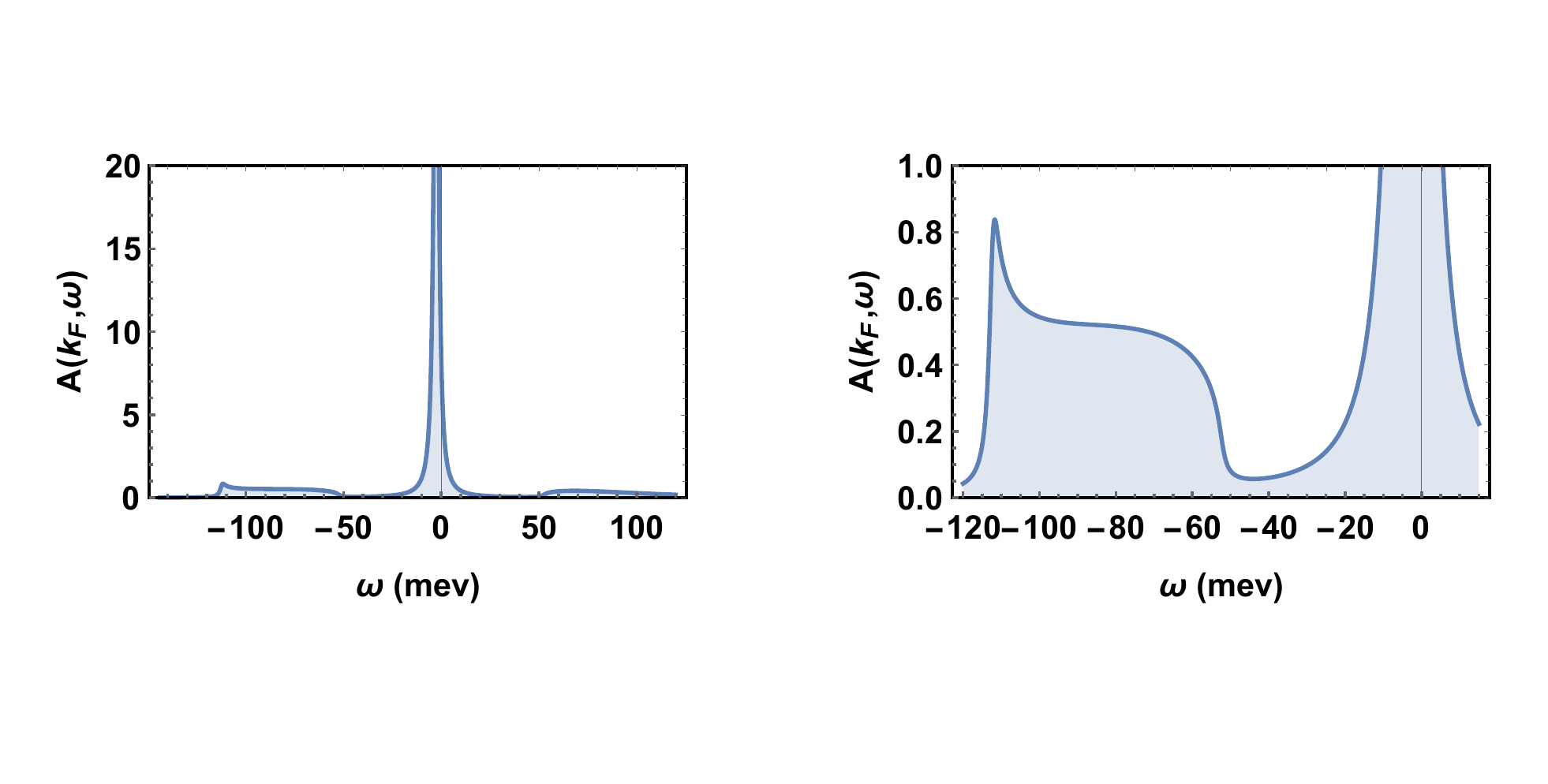}
\caption{ Spectral weight of electron - like excitation. Left: overview of
the main quasiparticle and two satellites (ARPES and inverse ARPES).}
\end{figure}

\subsection{The shape of the quasiparticle satellites}

The shape of the spectral weight $A\left( \mathbf{k},\omega \right) $ at $%
\left \vert \mathbf{k}\right \vert =k_{F}$ was calculated for $\Omega
=\Omega _{s}$ (see Fig.12, left panel). One observes that beyond the
dominant sharp quasiparticle peak near $\omega =0$, there are two small
"satellite" structures created by the soft phonon mode. The one with the
spectral weight of $0.13$, observed in ARPES \cite{Lee12,isotopeGuo},
extends (see Fig. 12 right panel) from the phonon mode $\omega =-\Omega _{s}$
all the way to the peak at the satellite location slightly above $\omega
=-\Omega _{s}-E_{F}/\hbar \approx -100mev$. The satellite excitation,
associated with the hard mode $\Omega _{h}$, would appear at much lower
energies and with lower weight.

\subsection{Dimensionless electron - electron coupling $\protect \lambda $}

The coupling constant is defined in terms of the self energy analytically
continued to the physical frequencies in the limit $\omega \rightarrow 0$) 
\begin{equation}
\lambda =-\frac{d}{d\omega }\Sigma \left( k_{F},\omega \right) |_{\omega =0}=%
\frac{\left( Z_{O}e^{2}\rho \right) ^{2}}{8\pi ^{2}M\Omega }%
\int_{l=0}^{2k_{F}}le^{-2l/q_{0}}I_{\lambda }\left( l\right)
\label{4-lamdef}
\end{equation}%
One again accounts for the step function $\Theta \left[ -\xi _{l,\phi }%
\right] $ function as $l+2k_{F}\cos \phi <0$, leading to the limiting value
of $\phi _{0}=\arccos \left[ -l/2k_{F}\right] $:

\begin{eqnarray}
I_{\lambda }\left( l\right) &=&\int_{\phi =0}^{2\pi }\left( \frac{\Theta %
\left[ -\xi _{l,\phi }\right] }{\left( \xi _{l,\phi }-\Omega \right) ^{2}}+%
\frac{\Theta \left[ \xi _{l,\phi }\right] }{\left( \xi _{l,\phi }+\Omega
\right) ^{2}}\right)  \label{4_Idef} \\
&=&\int_{\phi =\phi _{0}}^{\pi }\frac{2}{\left( \frac{l+2k_{F}\cos \phi }{%
2m^{\ast }}l-\Omega \right) ^{2}}+\int_{\phi =0}^{\phi _{0}}\frac{2}{\left( 
\frac{l+2k_{F}\cos \phi }{2m^{\ast }}l+\Omega \right) ^{2}}\text{.}  \notag
\end{eqnarray}%
It is important to perform the angle exactly in terms of analytic functions $%
f_{\lambda }^{1}$, $f_{\lambda }^{2}$ that are somewhat cumbersome. Direct
numerical integration suffers from extreme sensitivity near the Fermi level.
Changing the variable again to dimensionless $\overline{l}%
=l/2k_{F}\rightarrow l$, and $\widetilde{\Omega }=\frac{m^{\ast }}{2k_{F}^{2}%
}\Omega $ one writes, 
\begin{equation}
I_{\lambda }\left( l\right) =\frac{m^{\ast 2}}{2k_{F}^{4}l^{2}}\left(
f_{\lambda }^{1}\left[ l-\widetilde{\Omega }/l\right] +f_{\lambda }^{2}\left[
-l-\widetilde{\Omega }/l\right] \right) \text{,}  \label{4_Ilambda}
\end{equation}%
where

\begin{eqnarray}
f_{\lambda }^{1}\left[ a\right] &=&\frac{1}{a^{2}-1}\left \{ \frac{\sqrt{%
1-l^{2}}}{a+l}+\frac{a}{\sqrt{\left \vert a^{2}-1\right \vert }}R^{1}\left[ a%
\right] \right \}  \label{4_int} \\
f_{\lambda }^{1,2}\left[ a\right] &=&\frac{1}{a^{2}-1}\left \{ \frac{\sqrt{%
1-l^{2}}}{a\mp l}+\frac{a}{\sqrt{\left \vert a^{2}-1\right \vert }}R^{1,2}%
\left[ a\right] \right \}  \notag \\
R^{1}\left[ a\right] &=&\left \{ 
\begin{array}{c}
2\  \text{arccot}\left[ \frac{\sqrt{a^{2}-1}}{a-1}\tan \frac{\phi _{0}}{2}%
\right] \text{ for }a<-1 \\ 
\text{Re log}\frac{\sqrt{1-a^{2}}\tan \left[ \phi _{0}/2\right] +a+1}{\sqrt{%
1-a^{2}}\tan \left[ \phi _{0}/2\right] -a-1}\text{ for }-1<a<-l%
\end{array}%
\right \vert ;  \notag \\
R^{2}\left[ a\right] &=&\left \{ 
\begin{array}{c}
2\text{ arctan}\left[ \frac{\sqrt{a^{2}-1}}{a-1}\tan \frac{\phi _{0}}{2}%
\right] \text{ for }a<-1 \\ 
-\text{Re log}\frac{\sqrt{1-a^{2}}\tan \left[ \phi _{0}/2\right] +a-1}{\sqrt{%
1-a^{2}}\tan \left[ \phi _{0}/2\right] -a+1}\text{ for }-1<a<-l%
\end{array}%
\right \vert .  \notag
\end{eqnarray}%
The dimensionless coupling constant therefore becomes 
\begin{equation}
\lambda =\frac{g^{2}k_{F}^{2}}{\widetilde{\Omega }}\int_{l=0}^{1}\frac{%
e^{-4k_{F}l/q_{c}}}{l}\left( f_{\lambda }^{1}\left[ l-\widetilde{\Omega }/l%
\right] +f_{\lambda }^{2}\left[ -l-\widetilde{\Omega }/l\right] \right) 
\text{,}  \label{4_lambda}
\end{equation}%
where the electron - phonon coupling definition, Eq.(\ref{4_dimensionless})
was used. This is convergent (the term in brackets is proportional to $%
\widetilde{l}$ at small $\widetilde{l}$) and was calculated numerically. \
The standard dimensionless electron phonon coupling\textit{\ }is from Eq. (%
\ref{4_Ilambda}) for the soft and hard modes are $\lambda _{s}=0.23$ and $%
\lambda _{h}=0.07$\ respectively. The first is larger than estimated from
the satellite experiments\cite{isotopeGuo}, while the second is smaller.
However the theoretical formula used in the estimate\cite%
{Lee12,JohnsonNJP16,Kulic} was derived on an assumption of \textit{delta -\
like} forward scattering peak for the hard mode. The soft mode value alone
would not be sufficient, if the BCS formula is applied: $T_{c}=1.14\Omega
e^{-1/\lambda _{s}}=9K$. Higher $T_{c}$ value above is caused by the forward
peak that is however just exponential, see Eq.(\ref{rho}), much wider than
conjectured delta function assumed in ref.\cite{Lee12,JohnsonNJP16,Kulic}.

\section{Discussion and conclusions.}

\textit{\ }To summarize, using a microscopic model of the ionic lattice
vibrations in the STO substrate below one unit cell $FeSe$, an "additional" $%
\Omega _{s}=50mev$ LO interface mode is identified, see Fig.6. The soft mode
propagating mainly in the first $TiO_{2}$ layer ("$O$ chains") has stronger
electron - phonon coupling to electron gas in $FeSe$ than a well known $%
\Omega _{h}=100mev$ hard mode. The increase seem to be solely due to reduced
frequency since the matrix elements of the electron - phonon interactions 
\cite{Grimvall,Eliashberg} are very similar for the two modes (numerous
other phonon modes \cite{DFT16,DFTnew} have significantly lower matrix
elements).

The coupling constant, critical temperature, replica band are calculated.
The numerical solution of the gap equations (as well as the Eliashberg
approximation to it) results in the $T_{c}=51K$ (while for the $^{18}O$
isotope it becomes $T_{c}=49K$). This result is both due to the reduced
phonon frequency and due to the spatial separation between the two
dimensional electron gas in the $FeSe$ layer and vibrating ions. The later
manifests itself in an exponential forward peak in the electron - phonon
scattering. It leads to a deviation from the BCS dependence of critical
temperature on $\lambda $. The coupling constant, $\lambda =0.23$, is strong
enough \textit{in this case} to account for most if not all of the huge
enhancement of the superconductivity on the $STO$ substrate compared to
parent compound $FeSe$. The peak is clearly not as sharp as assumed in
recent theories \cite{Lee12,JohnsonNJP16,Kulic}.

As to remarkable normal state properties of the 1UC $FeSe/STO$, the results
are following. The violation of the Migdal theorem is confirmed and
satellite excitations due to phonons appear in the spectral weight appear,
Fig. 12. The satellite is broad, but unlike in the delta function scattering
peak theory \cite{Lee12,JohnsonNJP16,Kulic} its divergence appears at
frequency much higher than $\Omega _{s}$ consistent with observations. We
discuss next possible signatures of the soft mode and generalizations of the
mechanism to other high $T_{c}$ materials.

The transversal (TO) counterpart of the LO soft mode considered here
indicates a close proximity of the ferroelectric instability of the
displacement type due to oxygen "empty site", see Fig. 5. Can this be
related to known phonon characteristics? Of course $STO$ is a perovskite
with very high dielectric constant "close" to ferroelectric transition\cite%
{Mahan}. First the soft surface mode considered here is not related to the
displacive structural transition\cite{Mahan,Petzelt} in bulk $STO$ at $105K$
(so called $A_{1g}$ mode has large frequencies at low temperature at become
soft at $105K$). There exists however another bulk $TO$ mode\cite%
{Mahan,Petzelt} ($E_{u}+A_{2u}$), that might be associated with the surface
soft mode. Its frequency strongly decreases with temperature and it
contributes to the large dielectric constant. Numerous surface measurements%
\cite{Xue16phonon,phonon} and density functional calculations\cite%
{DFT16,DFTnew} of phonons in the 1UC $FeSe$/$STO$ system indicate that there
are a few possible candidates in the relevant energy domain.

The present approach is a phenomenological in the sense that instead of
directly relying on the DFT simulations results for the phonon spectrum, one
utilizes the DFT results for the charge distributions in conjunction with
the experimental direct studies of the crystalline structure (greatly
enhances recently in view of progress in the STM and X rays techniques) in
the strongly ionic layers adjacent to 2D electron gas to infer about both
the dispersion of the relevant phonon modes and their coupling to charged
layer. These are factors that directly affects Cooper pairing. The explicit
identification of the dominant degrees of freedom is necessary for a
qualitative understanding of the pairing mechanism without the background of
plethora of other modes that exist in both the $FeSe$ unit cell and the
substrate material. Note that, unlike in other approaches, semi -
macroscopic quantities like dielectric constants are included on the
microscopic level.

Similar soft modes might exist in other high $T_{c}$ superconductors. For
example recently fabricated ultra - thin $CuO_{2}$ films on the $BiO$
surfaces of the $Bi-2212$ crystals were shown\cite{XueBSCCO} to exhibits
large s-wave gap in the $CuO_{2}$ layers. This perovskite allows the
microscopic approach outlined in the present work.

\textit{Acknowledgements. }

We are grateful Prof. Y. Guo, D. Li and L. L.Wang for helpful discussions.
Work of B.R. was supported by NSC of R.O.C. Grants No. 98-2112-M-009-014-MY3.

\section{Appendix A. The $TiO_{2}$ oxygen vibration modes}

The dominant degree of freedom (oxygen atoms in the interface $TiO_{2}$ on
two sublattices directly beneath the $4d$ $Fe$ orbitals) were described in
the text. The vibrations along the $z$ direction is also safely neglected.
The Hamiltonian for these degrees of freedom is%
\begin{equation}
H_{ph}=K_{ph}+W\text{,}  \label{2_Hph}
\end{equation}%
where kinetic energy is%
\begin{equation}
K_{ph}=\frac{M}{2}\sum \nolimits_{\mathbf{n}}\left \{ \left( \frac{d}{dt}%
\mathbf{u}_{\mathbf{n}}^{1}\right) ^{2}+\left( \frac{d}{dt}\mathbf{u}_{%
\mathbf{n}}^{2}\right) ^{2}\right \} \text{,}  \label{2_Kph}
\end{equation}%
and the potential energy part consists of interatomic potentials defined in
Eq.(\ref{interatomic}) and Table 1. Only interactions of the "dynamic"
oxygen atoms in the $TiO_{2}$ with neighboring $SrO$ below and $Se$ above
are taken into account:%
\begin{eqnarray}
W &=&\frac{1}{2}\sum \nolimits_{\mathbf{n,m},A}\left \{ v^{TiO}\left[ -%
\mathbf{r}_{\mathbf{m}}^{A}-\mathbf{u}_{\mathbf{m}}^{A}\right] +v^{SrO}\left[
\mathbf{R}_{\mathbf{n}}^{Sr}-\mathbf{r}_{\mathbf{m}}^{A}-\mathbf{u}_{\mathbf{%
m}}^{A}\right] +v^{SeO}\left[ \mathbf{R}_{\mathbf{n}}^{Se}-\mathbf{r}_{%
\mathbf{m}}^{A}-\mathbf{u}_{\mathbf{m}}^{A}\right] +v^{OO}\left[ \mathbf{R}_{%
\mathbf{n}}^{O}-\mathbf{r}_{\mathbf{m}}^{A}-\mathbf{u}_{\mathbf{m}}^{A}%
\right] \right \}  \label{2_potW} \\
&&+\frac{1}{2}\sum \nolimits_{\mathbf{n\not=m},A}v^{OO}\left[ \mathbf{r}_{%
\mathbf{n}}^{A}-\mathbf{r}_{\mathbf{m}}^{A}+\mathbf{u}_{\mathbf{n}}^{A}-%
\mathbf{u}_{\mathbf{m}}^{A}\right] +\sum \nolimits_{\mathbf{n,m}}v^{OO}\left[
\mathbf{r}_{\mathbf{n}}^{1}-\mathbf{r}_{\mathbf{m}}^{2}+\mathbf{u}_{\mathbf{n%
}}^{1}-\mathbf{u}_{\mathbf{m}}^{2}\right] \text{.}  \notag
\end{eqnarray}%
Here the positions of the heavy $Ti,Sr,Se$ atoms and oxygen atoms of the $%
SrO $ layer are, 
\begin{eqnarray}
\mathbf{R}_{\mathbf{n}}^{Ti} &=&a\left( n_{x},n_{y},0\right) ;  \label{2-pos}
\\
\mathbf{R}_{\mathbf{n}}^{Sr} &=&\mathbf{R}_{\mathbf{n}}^{Se}=a\left( n_{x}+%
\frac{1}{2},n_{y}+\frac{1}{2},z_{Sr}\right) ;  \notag \\
\mathbf{R}_{\mathbf{n}}^{O} &=&a\left( n_{x},n_{y},z_{Sr}\right) \text{,} 
\notag
\end{eqnarray}%
see Figs. 1-3. Vibrations of heavy atoms and even oxygen in other planes are
not expected to be significant due to their mass or distance from the $TiO$
layer oxygen atoms. Some effects of those vibrations is accounted for by the
effective oxygen mass, while more remote $Fe$ later above and next $TiO_{2}$
below the important layer were checked to be negligible.

Harmonic approximation consists of expansion around a stable minimum of the
energy. The matrix of the second derivatives include:

\begin{eqnarray}
\frac{d^{2}W}{du_{\mathbf{m}}^{\alpha 1}du_{\mathbf{l}}^{\beta 1}}
&=&-\delta _{\mathbf{ml}}\sum \nolimits_{\mathbf{n}}\left \{ v_{\alpha \beta
}^{TiO}\left[ \mathbf{R}_{\mathbf{n}}^{Ti}-\mathbf{r}_{\mathbf{m}}^{1}\right]
+v_{\alpha \beta }^{SrO}\left[ \mathbf{R}_{\mathbf{n}}^{Sr}-\mathbf{r}_{%
\mathbf{m}}^{1}\right] +v_{\alpha \beta }^{SeO}\left[ \mathbf{R}_{\mathbf{n}%
}^{Se}-\mathbf{r}_{\mathbf{m}}^{1}\right] +v_{\alpha \beta }^{OO}\left[ 
\mathbf{R}_{\mathbf{n}}^{O}-\mathbf{r}_{\mathbf{m}}^{1}\right] \right \}
\label{1_second} \\
&&+\delta _{\mathbf{ml}}\sum \nolimits_{\mathbf{n\not=m}}v_{\alpha \beta
}^{OO}\left[ \mathbf{r}_{\mathbf{m}}^{1}-\mathbf{r}_{\mathbf{n}}^{1}\right]
-v_{\alpha \beta }^{OO}\left[ \mathbf{r}_{\mathbf{m}}^{1}-\mathbf{r}_{%
\mathbf{l}}^{1}\right] \sum \nolimits_{\mathbf{n\not=m}}\delta _{\mathbf{nl}%
}+\delta _{\mathbf{ml}}\sum \nolimits_{\mathbf{n}}v_{\alpha \beta }^{OO}%
\left[ \mathbf{r}_{\mathbf{m}}^{1}-\mathbf{r}_{\mathbf{n}}^{2}\right] ; 
\notag \\
\frac{d^{2}W}{du_{\mathbf{m}}^{\alpha 1}du_{\mathbf{l}}^{\beta 2}}
&=&-v_{\alpha \beta }^{OO}\left[ \mathbf{r}_{\mathbf{m}}^{1}-\mathbf{r}_{%
\mathbf{l}}^{2}\right] ;\text{ \ }\frac{d^{2}W}{du_{\mathbf{m}}^{\alpha
2}du_{\mathbf{l}}^{\beta 1}}=-v_{\alpha \beta }^{OO}\left[ \mathbf{r}_{%
\mathbf{m}}^{2}-\mathbf{r}_{\mathbf{l}}^{1}\right] ;  \notag \\
\frac{d^{2}W}{du_{\mathbf{m}}^{\alpha 2}du_{\mathbf{l}}^{\beta 2}}
&=&-\delta _{\mathbf{ml}}\sum \nolimits_{\mathbf{n}}\left \{ v_{\alpha \beta
}^{TiO}\left[ \mathbf{R}_{\mathbf{n}}^{Ti}-\mathbf{r}_{\mathbf{m}}^{2}\right]
+v_{\alpha \beta }^{SrO}\left[ \mathbf{R}_{\mathbf{n}}^{Sr}-\mathbf{r}_{%
\mathbf{m}}^{2}\right] +v_{\alpha \beta }^{SeO}\left[ \mathbf{R}_{\mathbf{n}%
}^{Se}-\mathbf{r}_{\mathbf{m}}^{2}\right] +v_{\alpha \beta }^{OO}\left[ 
\mathbf{R}_{\mathbf{n}}^{O}-\mathbf{r}_{\mathbf{m}}^{2}\right] \right \} 
\notag \\
&&+\delta _{\mathbf{ml}}\sum \nolimits_{\mathbf{n\not=m}}v_{\alpha \beta
}^{OO}\left[ \mathbf{r}_{\mathbf{m}}^{2}-\mathbf{r}_{\mathbf{n}}^{2}\right]
+v_{\alpha \beta }\left[ \mathbf{r}_{\mathbf{m}}^{2}-\mathbf{r}_{\mathbf{l}%
}^{2}\right] \sum \nolimits_{\mathbf{n\not=m}}\delta _{\mathbf{nl}}+\delta _{%
\mathbf{ml}}\sum \nolimits_{\mathbf{n}}v_{\alpha \beta }^{OO}\left[ \mathbf{r%
}_{\mathbf{n}}^{1}-\mathbf{r}_{\mathbf{m}}^{2}\right] \text{.}  \notag
\end{eqnarray}%
Here

\begin{equation}
v_{\alpha \beta }^{XY}\left[ \mathbf{r}\right] \equiv \frac{d^{2}v^{XY}}{%
dr_{\alpha }dr_{\beta }}=\frac{e^{2}Z_{X}Z_{Y}}{r^{5}}\left( 3r^{\alpha
}r^{\beta }-\delta ^{\alpha \beta }r^{2}\right) +\frac{\sqrt{A_{X}A_{Y}}b}{%
r^{3}}\left \{ -\delta _{\alpha \beta }r^{2}+r_{\alpha }r_{\beta }\left(
1+br\right) \right \} e^{-br}\text{,}  \label{1_v}
\end{equation}%
with $b\equiv \frac{1}{2}\left( b^{X}+b^{Y}\right) $.

Fourier transform defined as\ 

\begin{equation}
\mathbf{u}_{\mathbf{k}}^{A}=\frac{1}{N_{s}}\sum \nolimits_{\mathbf{m}}exp%
\left[ -\frac{2\pi yi}{N_{s}}\mathbf{k\cdot m}\right] \mathbf{u}_{\mathbf{m}%
}^{A}\text{,}  \label{2_Fourier_u}
\end{equation}%
where $N_{s}^{2}$ is the number of unit cells. This leads to the following
expression for the dynamic matrix

\begin{eqnarray}
D_{\mathbf{k}}^{\alpha 1\beta 2} &=&-\frac{1}{N_{s}}\sum \nolimits_{\mathbf{n%
}}\exp \left[ -\frac{2\pi i}{N_{s}}\mathbf{k\cdot n}\right] v_{\alpha \beta
}^{OO}\left[ \mathbf{r}_{\mathbf{n}}^{1}-\mathbf{r}_{\mathbf{0}}^{2}\right] ;
\label{2_D} \\
D_{\mathbf{k}}^{\alpha 2\beta 1} &=&-\frac{1}{N_{s}}\sum \nolimits_{\mathbf{n%
}}\exp \left[ -\frac{2\pi i}{N_{s}}\mathbf{k\cdot n}\right] v_{\alpha \beta
}^{OO}\left[ \mathbf{r}_{\mathbf{0}}^{1}-\mathbf{r}_{\mathbf{n}}^{2}\right] ;
\notag \\
D_{\mathbf{k}}^{\alpha 1\beta 1} &=&D_{\mathbf{k}}^{\alpha 2\beta 2}=\frac{1%
}{N_{s}}\left \{ 
\begin{array}{c}
\sum \nolimits_{\mathbf{n}}v_{\alpha \beta }^{OO}\left[ \mathbf{r}_{\mathbf{n%
}}^{1}-\mathbf{r}_{\mathbf{0}}^{2}\right] \\ 
-\sum \nolimits_{\mathbf{n}}\left \{ v_{\alpha \beta }^{TiO}\left[ \mathbf{R}%
_{\mathbf{n}}^{Ti}-\mathbf{r}_{\mathbf{0}}^{2}\right] +v_{\alpha \beta
}^{SrO}\left[ \mathbf{R}_{\mathbf{n}}^{Sr}-\mathbf{r}_{\mathbf{0}}^{2}\right]
+v_{\alpha \beta }^{SeO}\left[ \mathbf{R}_{\mathbf{n}}^{Se}-\mathbf{r}_{%
\mathbf{0}}^{2}\right] +v_{\alpha \beta }^{OO}\left[ \mathbf{R}_{\mathbf{n}%
}^{O}-\mathbf{r}_{\mathbf{0}}^{2}\right] \right \} \\ 
+\sum \nolimits_{\mathbf{n\not=0}}\left( 1-\exp \left[ -\frac{2\pi i}{N_{s}}%
\mathbf{k\cdot n}\right] \right) v_{\alpha \beta }^{OO}\left[ \mathbf{r}_{%
\mathbf{n}}^{2}-\mathbf{r}_{\mathbf{0}}^{2}\right]%
\end{array}%
\right \} \text{.}  \notag
\end{eqnarray}%
These determine the eigenvalues and polarizations presented in Figs. 5 and
Fig. 6 respectively.

\section{Appendix B. Derivation of Gorkov equations for a two band system}

We derive the Gorkov's equations within the functional integral approach\cite%
{NO,frontiers} starting from the effective electron action Eqs.(9),(13) for
grassmanian fields $\psi _{\mathbf{k},n}^{\ast \sigma A}$ and $\psi _{%
\mathbf{k},n}^{\sigma A}$:

\begin{equation}
\mathcal{A}\left[ \psi \right] =T^{-1}\sum \nolimits_{\mathbf{k}n}\psi _{%
\mathbf{k}n}^{\ast \sigma A}\left( G_{\mathbf{k}n}^{0}\right) ^{-1}\psi _{%
\mathbf{k}n}^{\sigma A}+\frac{1}{2T}\sum \nolimits_{\mathbf{q}.n}n_{\mathbf{q%
}n}^{Y}v_{\mathbf{q}n}^{YX}n_{-\mathbf{q,}n}^{X}\text{.}  \label{4_action}
\end{equation}%
To simplify the presentation it is useful to lump the quasi - momentum and
the Matsubara frequency into a single subscript, $\left \{ \mathbf{k}%
,n\right \} \rightarrow \kappa $. In this form (all the repeated indices are
assumed to be summed over), the action is:

\begin{equation}
\mathcal{A}\left[ \psi \right] =T^{-1}\psi _{\alpha }^{\ast \sigma A}\left(
G_{\alpha }^{0}\right) ^{-1}\psi _{\alpha }^{\sigma A}+\frac{1}{2T}\psi
_{\beta }^{\ast \sigma Y}\psi _{\chi +\beta }^{\sigma Y}v_{\chi }^{YX}\psi
_{\gamma }^{\ast \rho X}\psi _{\gamma -\chi }^{\rho X}\text{.}
\label{5_Aeff}
\end{equation}%
Functional derivative of the partition sum, 
\begin{equation}
Z=\int_{\psi }e^{-\mathcal{A}\left[ \psi ^{\ast },\psi \right] -J_{\alpha
}^{\ast \sigma A}\psi _{\alpha }^{\sigma A}-\psi _{\alpha }^{\ast \sigma
A}J_{\alpha }^{\sigma A}}\text{,}  \label{4_Z}
\end{equation}%
to the following gaussian average of the "equations of state",

\begin{eqnarray}
J_{\beta }^{\sigma B} &=&-\left \langle \frac{\delta \mathcal{A}}{\delta
\psi _{\beta }^{\ast \sigma B}}\right \rangle =-\left( G_{\beta }^{0}\right)
^{-1}\psi _{\beta }^{\sigma B}+v_{\chi }^{BX}\left \langle \psi _{\gamma
}^{\ast \rho X}\psi _{\chi +\beta }^{\sigma B}\right \rangle \psi _{\gamma
-\chi }^{\rho X}  \label{4_J} \\
&&+v_{\chi }^{BX}\left \langle \psi _{\chi +\beta }^{\sigma B}\psi _{\gamma
-\chi }^{\rho X}\right \rangle \psi _{\gamma }^{\ast \rho X}-v_{\chi
}^{BX}\left \langle \psi _{\gamma }^{\ast \rho X}\psi _{\gamma -\chi }^{\rho
X}\right \rangle \psi _{\chi +\beta }^{\sigma B}\text{.}  \notag
\end{eqnarray}%
Translation invariance and the s-wave Ansatz,

\begin{eqnarray}
\left \langle \psi _{\mathbf{\beta }}^{\rho B}\psi _{\mathbf{\alpha }}^{\ast
\sigma A}\right \rangle &=&\delta _{\alpha -\beta }\delta ^{\sigma \rho
}G_{\alpha }^{AB};  \label{4_Ansatz} \\
\left \langle \psi _{\mathbf{\alpha }}^{\sigma A}\psi _{\beta }^{\rho
B}\right \rangle &=&\delta _{\alpha +\beta }\varepsilon ^{\sigma \rho
}F_{\alpha }^{AB};  \notag \\
\left \langle \psi _{\mathbf{\alpha }}^{\ast \sigma A}\psi _{\beta }^{\ast
\rho B}\right \rangle &=&\delta _{\alpha +\beta }\varepsilon ^{\sigma \rho
}F_{\alpha }^{+AB}\text{,}  \notag
\end{eqnarray}%
lead lead to:%
\begin{equation}
J_{\beta }^{\sigma B}=-\left( G_{\beta }^{0}\right) ^{-1}\psi _{\beta
}^{\sigma B}-v_{\chi }^{BX}G_{\chi +\beta }^{XB}\psi _{\beta }^{\sigma
X}+\varepsilon ^{\sigma \rho }v_{\chi }^{BX}F_{\chi +\beta }^{BX}\psi _{%
\mathbf{-}\beta }^{\ast \rho X}+2v_{0}^{BX}\psi _{\beta }^{\sigma B}G_{\chi
}^{XX}\text{.}  \label{4_Jeq1}
\end{equation}%
Similarly

\begin{eqnarray}
J_{\beta }^{\ast \sigma B} &=&\left \langle \frac{\delta \mathcal{A}}{\delta
\psi _{\beta }^{\sigma B}}\right \rangle =-\left( G_{\beta }^{0}\right)
^{-1}\psi _{\beta }^{\ast \sigma B}-v_{\chi }^{BX}\left \langle \psi _{\beta
-\chi }^{\ast \sigma B}\psi _{\gamma }^{\ast \rho X}\right \rangle \psi
_{\gamma -\chi }^{\rho X}  \label{4_Jstareq} \\
&&+v_{\chi }^{BX}\left \langle \psi _{\beta -\chi }^{\ast \sigma B}\psi
_{-\chi +\gamma }^{\rho X}\right \rangle \psi _{\gamma }^{\ast \rho
X}-v_{\chi }^{BX}\psi _{\beta -\chi }^{\ast \sigma B}\left \langle \psi
_{\gamma }^{\ast \rho X}\psi _{-\chi +\gamma }^{\rho X}\right \rangle  \notag
\\
&=&\left( G_{\beta }^{0}\right) ^{-1}\psi _{\beta }^{\ast \sigma
B}-\varepsilon ^{\sigma \rho }v_{\chi }^{BX}F_{\beta -\chi }^{+BX}\psi
_{-\beta }^{\rho X}-v_{\chi }^{BX}G_{\beta -\chi }^{XB}\psi _{\beta }^{\ast
\sigma X}+2v_{0}^{BX}\psi _{\beta }^{\ast \sigma B}G_{\chi }^{XX}\text{.} 
\notag
\end{eqnarray}

The second derivatives with respect to fields are$_{\text{,}}$%
\begin{eqnarray}
\frac{\delta J_{\beta }^{\ast \sigma B}}{\delta \psi _{\gamma }^{\rho C}}
&=&-\delta _{\beta +\gamma }\varepsilon ^{\sigma \rho }v_{\chi
}^{BC}F_{\beta -\chi }^{+BC};  \label{4_secder} \\
\frac{\delta J_{\beta }^{\ast \sigma B}}{\delta \psi _{\gamma }^{\ast \rho C}%
} &=&\delta _{\beta -\gamma }\delta ^{\sigma \rho }\left( \delta ^{BC}\left(
G_{\beta }^{0}\right) ^{-1}-v_{\chi }^{BC}G_{\beta -\chi }^{CB}-2\delta
^{BC}v_{0}^{BX}G_{\chi }^{XX}\right) ;  \notag \\
\frac{\delta J_{\beta }^{\sigma B}}{\delta \psi _{\gamma }^{\rho C}}
&=&\delta ^{\sigma \rho }\delta _{\beta -\gamma }\left( -\delta ^{BC}\left(
G_{\beta }^{0}\right) ^{-1}-v_{\chi }^{BC}G_{\chi +\beta }^{BC}+2\delta
^{BC}v_{0}^{BX}G_{\chi }^{XX}\right) ;  \notag \\
\frac{\delta J_{\beta }^{\sigma B}}{\delta \psi _{\gamma }^{\ast \rho C}}
&=&\varepsilon ^{\sigma \rho }\delta _{\beta +\gamma }v_{\chi }^{BC}F_{\chi
+\beta }^{BC}\text{.}  \notag
\end{eqnarray}

The Gorkov equations are obtained from the following identity

\begin{eqnarray}
\left \langle \psi _{\beta }^{\ast \sigma B}\psi _{\alpha }^{\theta A}\right
\rangle \frac{\delta J_{\beta }^{\sigma B}}{\delta \psi _{\gamma }^{\ast
\rho C}}+\left \langle \psi _{\alpha }^{\theta A}\psi _{\beta }^{\sigma
B}\right \rangle \frac{\delta J_{\beta }^{\ast \sigma B}}{\delta \psi
_{\gamma }^{\ast \rho C}} &=&0;  \label{4_identity} \\
\left \langle \psi _{\alpha }^{\theta A}\psi _{\beta }^{\ast \sigma B}\right
\rangle \frac{\delta J^{\sigma B}}{\delta \psi _{\gamma }^{\rho C}}+\left
\langle \psi _{\alpha }^{\theta A}\psi _{\beta }^{\sigma B}\right \rangle 
\frac{\delta J_{\beta }^{\ast \sigma B}}{\delta \psi ^{\rho C}} &=&\delta
^{\theta \rho }\delta ^{AC}\text{.}  \notag
\end{eqnarray}%
Substituting Eqs.(\ref{4_Ansatz}\ref{4_secder}), one finally obtains the
first,

\begin{equation}
G_{\alpha }^{CA}\left( G_{\alpha }^{0}\right) ^{-1}+\sum
\nolimits_{B}G_{\alpha }^{BA}v_{\chi }^{BC}G_{\chi +\alpha }^{BC}-2\sum
\nolimits_{X}G_{\alpha }^{CA}v_{0}^{CX}G_{\chi }^{XX}+\sum
\nolimits_{B}v_{\chi }^{BC}F_{\alpha }^{AB}F_{-\alpha -\chi }^{+BC}=\delta
^{AC}\text{,}  \label{4_Gor1}
\end{equation}%
and the second Gorkov equations,

\begin{equation}
F_{\alpha }^{AC}\left( G_{-\alpha }^{0}\right) ^{-1}+G_{\alpha }^{BA}v_{\chi
}^{BC}F_{\chi +\alpha }^{BC}+F_{\alpha }^{AB}v_{\chi }^{BC}G_{-\alpha -\chi
}^{CB}+2F_{\alpha }^{AC}\sum \nolimits_{X}v_{0}^{CX}G_{\chi }^{XX}=0\text{.}
\label{4_Gor2}
\end{equation}%
\newpage

The system of Gorkov equations, Eqs.(\ref{4_Gor1},\ref{4_Gor2}) simplifies
near the criticality. The last term in Eq.(\ref{4_Gor1}) is of order $F^{2}$
and thus negligible. The second and the third terms are small corrections to
the normal state Greens function at weak electron - phonon coupling.
Therefore one obtains from Eq.(\ref{4_Gor1})

\begin{equation}
G_{\alpha }^{AB}=\delta ^{AB}G_{\alpha }^{0}\text{.}  \label{4_Gor1a}
\end{equation}%
Substituting this into the second Gorkov equation, Eq.(\ref{4_Gor2}), one
obtains:%
\begin{equation}
F_{\alpha }^{AC}+K_{\alpha }^{C}v_{\chi }^{AC}F_{\chi +\alpha }^{AC}=0\text{,%
}  \label{4_Gor1b}
\end{equation}%
where%
\begin{equation}
K_{\alpha }^{C}=\frac{G_{-\alpha }^{0}G_{\alpha }^{0}}{1+G_{-\alpha
}^{0}v_{\chi }^{CC}G_{-\alpha -\chi }^{0}+2G_{-\alpha }^{0}\sum
\nolimits_{X}v_{0}^{CX}G_{\chi }^{0}}\text{.}  \label{4_Kdef}
\end{equation}%
In the denominator one argues that at weak coupling the first order
corrections can be neglected.

\bigskip

\section{Appendix C. Solution of the gap equation in the Eliashberg
approximation}

In polar coordinates for an angle independent gap function, $\Delta _{%
\mathbf{p,}m}^{AC}=\Delta _{k_{F}\mathbf{,}m}^{AC}\equiv \Delta _{m}^{AC}$,
and shifting the integration variables as the equation for momentum $\mathbf{%
q}$ on the Fermi surface, $\varepsilon _{\mathbf{q}}=\varepsilon _{F}$,
takes a form: 
\begin{equation}
-T_{c}\sum \nolimits_{m}\frac{1}{\left( 2\pi \right) ^{2}}\int_{\mathbf{l}}%
\frac{v_{\mathbf{l},n-m}^{AC}}{\left( \omega _{m}^{e}\right) ^{2}+\left(
\epsilon _{\left \vert \mathbf{q+l.}\right \vert }-\epsilon _{F}\right) ^{2}}%
\Delta _{m}^{AC}=\Delta _{n}^{AC}\text{.}  \label{gap}
\end{equation}%
The left hand side of the equation using the fit Eq.(\ref{rho}), can be
written in polar coordinates as 
\begin{equation}
\frac{T_{c}\left( Z^{O}e^{2}\right) ^{2}\rho ^{2}}{M\left( 2\pi \right) ^{2}}%
\sum \nolimits_{m}\frac{\Delta _{m}^{AC}}{\left( \omega _{n-m}^{b}\right)
^{2}+\Omega ^{2}}\int_{l=0}^{1/a}le^{-2l/q_{0}}\int_{\phi =0}^{2\pi }\frac{1%
}{\left( \omega _{m}^{f}\right) ^{2}+\xi _{l,\phi }^{2}}\text{,}
\label{RHgap}
\end{equation}

Rescaling, $\widetilde{l}=l/2k_{F}$, one obtains:%
\begin{equation}
\frac{T_{c}\left( Z_{O}e^{2}\rho m^{\ast }\right) ^{2}}{4\pi ^{2}Mk_{F}^{2}}%
\sum \nolimits_{m}\frac{\Delta _{m}^{AC}}{\left( \omega _{n-m}^{b}\right)
^{2}+\Omega ^{2}}\int_{\widetilde{l}=0}^{1}\frac{e^{-4k_{F}\overline{l}%
/q_{0}}}{\widetilde{l}}\int_{\phi =0}^{2\pi }\frac{1}{\left( m^{\ast }\omega
_{m}^{f}\right) ^{2}/\left( 2k_{F}^{2}\widetilde{l}\right) ^{2}+\left( 
\overline{l}+\cos \phi \right) ^{2}}\text{.}  \label{3_angint}
\end{equation}%
Integrating exactly over the angle $\phi $,

\begin{equation}
\frac{1}{2\pi }\int_{\phi =0}^{2\pi }\frac{1}{\left( s/l\right) ^{2}+\left(
l+\cos \phi \right) ^{2}}=\frac{l}{s}\text{Re}\left[ \left(
1+s^{2}/l^{2}-2is-l^{2}\right) ^{-1/2}\right] \text{,}  \label{3_equality}
\end{equation}%
one obtains, dropping tilde over $\widetilde{l}$ in what follows:

\begin{equation}
\frac{\left( Z^{O}e^{2}\rho \right) ^{2}m^{\ast }}{\pi ^{2}M}\sum
\nolimits_{m}\frac{\Delta _{m}^{AC}}{\left( \left( \omega _{n-m}^{b}\right)
^{2}+\Omega ^{2}\right) \left \vert 2m+1\right \vert }f\left[ \frac{\omega
_{m}^{f}m^{\ast }}{2k_{F}^{2}}\right] \text{.}  \label{3_angleint3}
\end{equation}%
Here the function is defined as an integral:

\begin{equation}
f\left[ z\right] =\int_{l=0}^{1}e^{-4k_{F}l/q_{0}}\text{Re}\left[ \frac{1}{%
\sqrt{1+z^{2}/l^{2}-2iz-l^{2}}}\right] \text{.}  \label{3-fdef}
\end{equation}%
The function $f\left[ z\right] $ and its rational fit $f\left[ z\right] =0.3%
\frac{1+z}{1+z+4z^{2}}$ are shown in Fig 13.

\begin{figure}[h]
\centering \includegraphics[width=10cm]{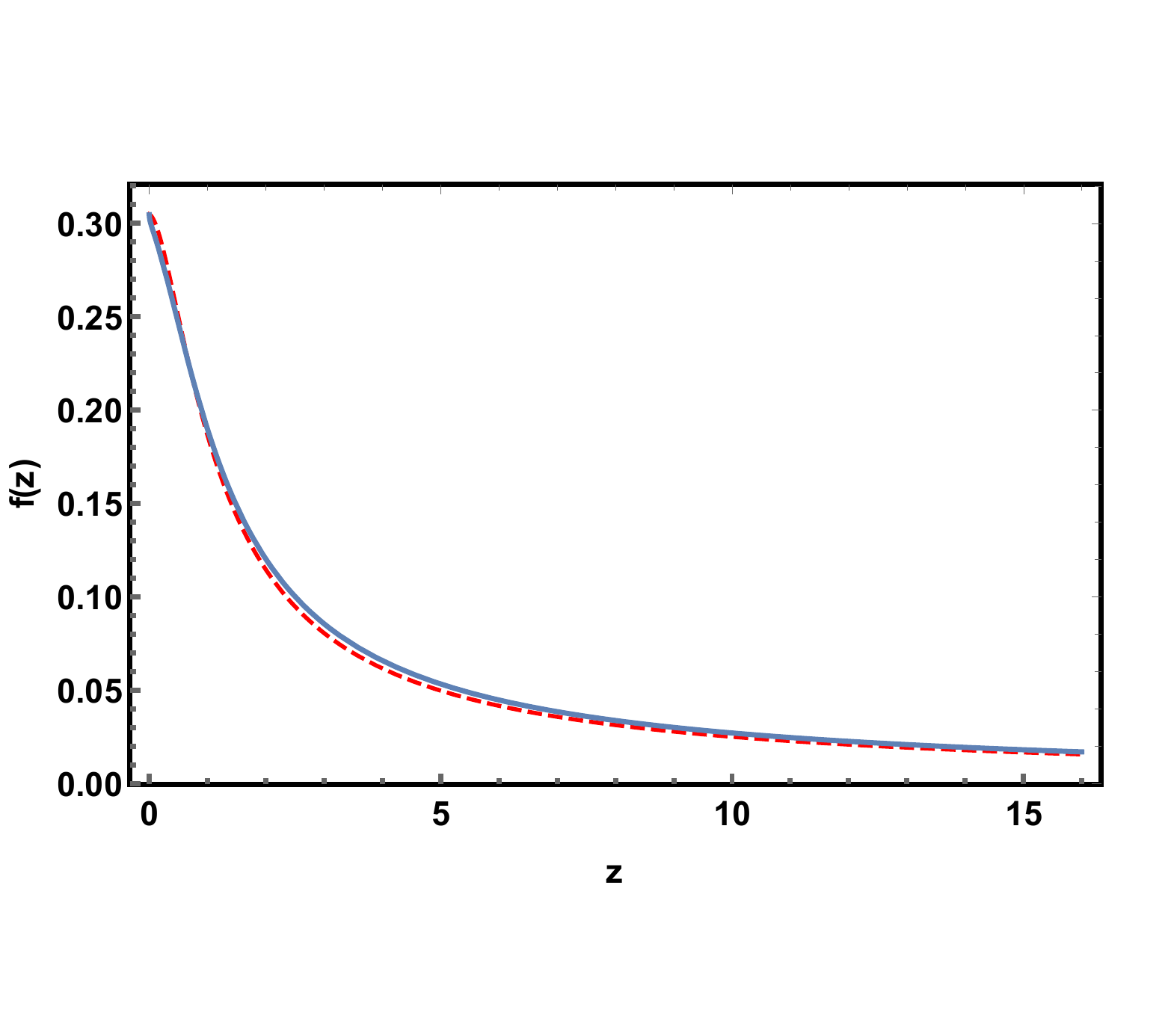}
\caption{Function $f\left[ z\right] $ in the gap equation, Eq.(22) in the
main text. }
\end{figure}

Changing the variables to $\eta _{n}=\sqrt{f\left( \omega _{n}^{f}m^{\ast
}/2k_{F}^{2}\right) /\left \vert \omega _{m}^{f}\right \vert }\Delta _{n}$,
makes the kernel matrix of the integral equation, 
\begin{equation}
\sum \nolimits_{m}K_{mn}\eta _{m}=\eta _{n}\text{,}  \label{4_eigeneq}
\end{equation}%
symmetric,%
\begin{equation}
K_{mn}=\frac{\left( \rho Z^{O}e^{2}\right) ^{2}m^{\ast }}{\pi ^{2}M\left(
\left( \omega _{n-m}^{b}\right) ^{2}+\Omega ^{2}\right) }\sqrt{\frac{f\left(
\omega _{m}^{f}m^{\ast }/2k_{F}^{2}\right) f\left( \omega _{n}^{f}m^{\ast
}/2k_{F}^{2}\right) }{\left \vert 2m+1\right \vert \left \vert 2n+1\right
\vert }}\text{.}  \label{4_Kphon}
\end{equation}%
Critical temperature is obtained when the largest eigenvalue of the matrix $%
K $ is unit. This was done numerically by limiting variable $n$ to $%
\left
\vert n\right \vert <200$.

Assuming as usual\cite{Eliashberg}, that the dependence of $\Delta $ on $%
\mathbf{k}$ is weak, $\Delta _{\mathbf{k,}n}^{AB}=\Delta _{n}^{AB}$,
substituting the soft mode $v_{\mathbf{l},n-m}^{AC}$ and integrating over
polar angle of $l$, the eigenvalue equation simplifies to

\begin{equation}
\frac{\left( Z_{O}e^{2}\rho \right) ^{2}m^{\ast }}{\pi ^{2}\Omega ^{2}M}\sum
\nolimits_{m}\frac{f\left[ \omega _{m}^{f}m^{\ast }/2k_{F}^{2}\right] }{%
\left( \left( \omega _{n-m}^{b}/\Omega \right) ^{2}+1\right) \left \vert
2m+1\right \vert }\Delta _{m}^{AC}=\Delta _{n}^{AC}\text{,}
\label{simplifiedgap}
\end{equation}%
Critical temperature is obtained when the largest eigenvalue of the matrix
in Eq.(\ref{simplifiedgap}) is unit. The numerical results are the
following. $T_{c}=51K$, while for isotope $^{18}O$ it becomes $T_{c}=49K$.

\end{document}